\documentclass[preprint,12pt]{elsarticle}

\usepackage{graphics}

\usepackage{amssymb, amsmath}
\usepackage{bm, bbm}

\biboptions{sort&compress}

\usepackage{physics}

\usepackage{hyperref}
\hypersetup{
    colorlinks=true,
    linkcolor=cyan,
    urlcolor=cyan,
}

\usepackage{listings}

\usepackage{blkarray}

 \graphicspath{{Figures/}}

\newcounter{bla}

\def\ie{\emph{i.e.} }

\journal{Computer Physics Communications}

\begin{document}

\begin{frontmatter}



\title{PULSEE: A software for the quantum simulation of an extensive set of   magnetic  resonance observables}


\author[a,c]{Davide Candoli}
\author[a]{Ilija K.\ Nikolov}
\author[a]{Lucas Z.\ Brito}
\author[a,b]{Stephen Carr}
\author[c]{Samuele Sanna}
\author[a]{Vesna F.  Mitrovi{\'c} \corref{VFM}}

 \cortext[VFM] {Vesna Mitrovi{\'c} \\\textit{E-mail address:} vemi@brown.edu}
\address[a]{Department of Physics, Brown University, Providence, 02912 Rhode Island, USA}
\address[b]{Brown Theoretical Physics Center, Brown University, Providence, Rhode Island 02912-1843, USA.}
\address[c]{Department of Physics and Astronomy ``A. Righi,'' University of Bologna, 40127 Bologna, Italy}

\begin{abstract}
We present an open-source software for the simulation of observables in magnetic  resonance experiments, including   nuclear magnetic/quadrupole resonance NMR/NQR and electron spin resonance (ESR)), developed to assist experimental research in the design of new strategies for the investigation of fundamental quantum properties of materials, as   inspired by  magnetic resonance protocols that emerged in the context of  quantum information science (QIS). 
The package introduced here enables  the simulation of both standard NMR spectroscopic observables and the time-evolution of an interacting single-spin system subject to complex pulse sequences, {\it i.e.} quantum gates. 
The main purpose of this software is to facilitate in the development of much needed novel NMR-based probes of emergent quantum orders, which can be elusive to standard experimental probes.  
The software is based on a  quantum mechanical description of nuclear spin dynamics in NMR/NQR experiments and has been widely tested on   available  theoretical and experimental  results. Moreover, the structure of the software allows for basic  experiments to  easily be  generalized to more sophisticated ones, as it includes all the libraries required for the numerical simulation of generic spin systems. In order to make the program easily accessible to a large user base, we  developed  a user-friendly graphical  interface,  Jupyter notebooks, and fully-detailed documentation.
Lastly, we portray several examples of the execution of the code  that illustrate   the potential of  a  novel NMR paradigm, inspired by QIS, for efficient investigation of emergent   phases in strongly correlated   materials.      
\end{abstract}


\end{frontmatter}



\noindent{\bf PROGRAM SUMMARY/NEW VERSION PROGRAM SUMMARY}

\begin{small}
\noindent
{\em Program Title: PULSEE (Program for the simULation of nuclear Spin Ensemble Evolution)}                                          \\
{\em CPC Library link to program files:} (to be added by Technical Editor) \\
{\em Developer's repository link:}  \url{https://github.com/vemiBGH/PULSEE} \\
{\em Code Ocean capsule:} (to be added by Technical Editor)\\
{\em Licensing provisions:} GPLv3 \\
{\em Programming language:} Python 3             \\                     
{\em Nature of problem:} 
Applications of nuclear magnetic/quadrupole resonance techniques to study the  properties of materials often requires extensive spectral simulations. On the other hand,  the application of   magnetic resonance techniques to quantum information science (QIS) involves different sets of observables. Available simulation software   address only one of these applications, that is, either detailed spectral simulations \cite{1} and/or QIS relevant observables \cite{2}. For this reason,  NMR has not seen as much development in the condensed matter community compared to  other spectroscopy techniques that combine these two approaches. 
Therefore, there is a need for an up-to-date and easily accessible software for the simulation of an extensive set of nuclear magnetic/quadrupole resonance experimental observables that allow  the behavior/response of nuclear systems with varying degree of complexity encountered in strongly correlated quantum materials to be  reproduced. \\
{\em Solution method:}  
 The open-source Python code provides an extensive set of libraries for the simulation of the time evolution of  spins in the presence of specific interactions and the reproduction of   spectra,  and other observables measured in  magnetic  resonance experiments,  as well as the simulations of quantum circuits and gates. The ready-to-use software features a user-friendly graphical  interface. \\

\end{small}

\section{Introduction}
\label{Intro}

Nuclear magnetic and quadrupole resonance (NMR/NQR) have a long-standing reputation as accurate methods for the microscopic investigation of materials based on remarkably simple working principles.  
In addition  to being a dominant tool in chemistry, materials science, structural biology,   and medicine, NMR represents an essential tool in quantum information science (QIS) \cite{RevModPhys_NMRQC, Ramanathan2004, Suter20QGate}. NMR can also be utilized for  fundamental tests   of quantum mechanics \cite{Modi12} and    condensed matter physics, as well as for probing microscopic spin and charge properties of materials \cite{Kaufmann79, Halperin86, Abragam, BlincIC}.  
These features are the reason for the success  of  magnetic resonance techniques    in   implementing   one of the first quantum information processors: the high degree of control of nuclear spins that they provide lends naturally to the purposes of basic quantum computing, and has made it possible to witness for the first time the experimental realization of several quantum algorithms \cite{Cory97, Chuang1998, Jones_1998, Jones_1998_2, Vandersypen_2000, LONG2001121, Sinha01, Havel02, Xin19}. Specifically, the handling of quantum systems to perform data processing tasks in NMR is accomplished through the application of specific radio frequency (RF) pulses (logic gates) on adequately prepared ensemble states, referred to pseudopure states (PPS) \cite{Sinha01, Havel02, NMRQIS_Book, Rao14, Teles18}. The logic gates can be executed with high fidelity due to  the superior level of control of the quantum evolution of nuclear spins. 
Indeed, a 12-qubit NMR based quantum computer   holds the record for the largest quantum computer, \ie high fidelity implementation of a quantum algorithm with coherent manipulation of 12 qubits \cite{Lu:2017aa}.   
Nonetheless, the long term   interest in the applications of NMR   in quantum computing has faded  since they present some major limitations when it comes to implementing a large scale quantum computer. 

Unfortunately in recent decades,  NMR has not seen as much development in the strongly correlated materials community compared to other spectroscopy techniques. The only place where NMR methodologies have kept on pace with our understanding of spin dynamics is as a control paradigm for quantum information technology (e.g. diamond-NV centers \cite{Liu2019}). Much of that progress has been in the realm of quantum control and sensing, {\it i.e.} the creation of specially engineered pulse sequences that best extract information out of single-spin systems \cite{RevModPhys.89.035002, PhysRevLett.102.210502,PhysRevX.8.021059, QMetrology20, Peng:2021vv}.  However, these protocols developed for the manipulation of  NMR qubits (single-spin systems) promise to be valuable resources for the exploration of  complex emergent  properties of  materials \cite{CarrMMSpec22,INikolov20}.  Here, we introduce unified protocols, presented in an open source software with a user-friendly interface,  to enable the simulation of both standard NMR spectroscopic observables and the time-evolution of an interacting single-spin system subject to complex pulse sequences, {\it i.e.} quantum gates. Our software can simulate the acquisition of the characteristic observable measured in a laboratory for single-spin systems under different pulse sequences, such as the free induction decay signal (FID), and then generate the NMR/NQR spectrum in a form which can be  directly compared with real experimental results.  The program is adaptable to the simulation of a wide range of experimental outcomes, as it makes use of  three different evolution solvers: {\it (i)} the time-independent Hamiltonian diagonal solver, {\it (ii)} the average Hamiltonian theory, 
implemented up to third order but easily extendable and {\it (iii)} QuTiP \cite{Johansson2012}. In addition, PULSEE   incorporates a quantum computing module that allows for the design of quantum circuit elements, relevant for both researchers and developers that focus on the direct, real-time  interface with instrumentation for quantum control, such as the Quantum Orchestration Platform  provided by Quantum Machines \cite{QuantumMachines}.

The main purpose of this software is to assist in the development of much needed novel NMR-based probes of emergent quantum orders, which can be elusive to standard experimental probes. Theoretically identified complex quantum phases of matter \cite{PhysRevLett.127.237201, Pourovskiie21} may encode details of their intricate structure into NMR responses  \cite{PhysRevLett.127.140604, Carr2021} in ways that lay outside the  current NMR spectroscopy paradigm. Therefore, our computational tool is instrumental in  designing the experiments (\ie NMR pulse sequences) that optimize the sensitivity of an NMR observable to the intricate structure of correlated quantum states of matter, as discussed in \mbox{Sec. \ref{HyperFSim}}. Moreover, the extension of this work to ensembles of nuclei will be vital for providing relevant data to enable the reverse engineering  of Hamiltonians of quantum phases of matter \cite{PhysRevLett.122.150606, PhysRevLett.124.100605}.  
 Finally, this program can be beneficial in designing optimal control protocols for quantum sensing applications.
    
Realizing  NMR protocols that ultimately enable the identification of quantum phases of matter   through the careful manipulation of nuclear spin degrees of freedom requires the development of   software to simulate experimental techniques, featuring the representation of nuclear spin states. Although there are many other NMR simulation programs, to our knowledge, most modern NMR/NQR software are mainly geared   towards applications in chemistry, or are add-on libraries to closed-source software. Some well liked, but aging programs that simulate NMR/NQR experiments are coded with less widely-used programming languages, such as SIMULDENS that uses VAS PASCAL \cite{ALLOUCHE1989171}, SIMPSON in the Tcl scripting language, while its core in the C programming language \cite{BAK2000296}, WSOLIDS1 in Microsoft Visual C++ 2008 Express Edition \cite{Eichele}, and WINDNMR-Pro, which is a stand-alone Windows programs whose development has ceased  and whose source code is not publicly available \cite{reich2002}. A similar simulation to ours is the NMR/NQR simulation that includes elliptically polarized RF fields with preparation of pseudo-pure states and basic quantum gates, proposed by Possa \textit{et al} \cite{Possa}, but its source is inaccessible, and it requires the paid Wolfram Mathematica environment. Other packages that are extensively used are \mbox{SpinDynamica \cite{Bengs2018}},  also for Mathematica, and Spinach for MATLAB \cite{Hogben2011}. There exist other licensed softwares, such as SpinEvolution~\cite{Veshtort2006} and the PERCH software, a wholly-owned subsidiary of Bruker BioSpin \cite{perchNMR}. Other programs that lack the density matrix visualization of spin states include QUEST \cite{PERRAS201236}  and SPINUS \cite{Binev2007}. For completeness, we note that numerous software packages have been developed in the computational chemistry community, but these are  mainly dedicated to   molecular  and protein structure determinations ~\cite{Claridge2009, Schwieters2001}. Therefore, our aim was to develop an up-to-date, open-source software, written in the more popular programming language Python, which   combines all the features relevant 
 to physics research, making them fully accessible and extensively documented. Furthermore, our software is integrated with the fairly known, and highly efficient QuTiP \cite{Johansson2012}, providing it with even more capabilities. The reason  
 for an initial   independent framework is to better understand and account for the technical difficulties, as opposed to using an existing framework as a black box. 

Our software PULSEE (Program for the simULation of nuclear Spin Ensemble Evolution) \cite{PULSEE} is based on the quantum mechanical description of magnetic resonance and is able to simulate the time evolution of nuclear spins   in a wide variety of configurations observed experimentally. The dynamics of the spin system is calculated in the interaction frame where the quantum states   only evolve  as a result of the time-dependent pulses, which makes the program highly versatile in its application. Although this package was designed to handle non-interacting, single-spin systems in solids dominated by the Zeeman and quadrupolar interactions, the software can handle relevant coupling with other nuclei and/or electrons, simulating the evolution of single-spin systems subject to different pulse sequences. As such, it is not intended for the direct reproduction of experiments that study   correlations and/or entanglement in quantum materials, but rather it allows for  the deviation of these experiments from an idealized single-spin evolution to be quantified. Once established, one may proceed to determine the source of the novel phenomena. To directly 
investigate strongly correlated phases of matter, one may use other  techniques and simulations of many-interacting spins.  
  In particular, one novel methodology allows the electronic susceptibility through  NMR to be   probed through the variation of  pulse strength and applied field orientation which  has direct applications for sensing and characterization of emergent electronic phases \cite{Carr2021, Snider2022}.  
  
The paper is organized as follows, in section \ref{ThBG} we give an overview of the theory of NMR and NQR, including both the description of nuclear spin dynamics and the generation of the spectra from the analysis of the FID. In section \ref{SUSoft} we present the simulation software, providing  practical information about its installation, structure, and usage. In section \ref{ExEx} we report several examples of simulations carried out with PULSEE, which have been chosen for their relevance to quantum control and quantum information processing. Specifically, in  \mbox{section \ref{HyperFSim}} we illustrate how ideas developed in the context of QIS can be deployed to efficiently probe the complexity of the hyperfine tensor arising as a result of intricate interactions in the emergent quantum phases of matter \cite{Lu17,RongOrbit19, PhysRevLett.127.237201, Pourovskiie21}.

\section{Theoretical background}
\label{ThBG}

Nuclear magnetic and quadrupole resonance (NMR/NQR) involve the time evolution of resonantly perturbed nuclear spins in matter. Experimentally, the distinction between the two methods lies in the different nuclear interactions being probed: NMR pertains to nuclei coupled to a local magnetic field (that is, an externally applied magnetic field), while NQR deals with the quadrupolar interaction between each nucleus and the surrounding electronic charges. From a theoretical point of view,  it is convenient to treat the problem where both interactions are simultaneously present, since it includes all the possible intermediate configurations between pure NMR and pure NQR. In addition, the system may include other less significant interactions that influence its evolution, such as dipole-dipole and hyperfine interactions, chemical and paramagnetic shift, $J$-coupling, and gradient field \cite{Abragam}. 

Although we aim to understand correlated systems, it is more beneficial to study single-spin, non-interacting systems, and gradually include interactions. The stationary Hamiltonian at thermal equilibrium is given by:
\begin{equation}
\mathcal{H}_{full} = \mathcal{H}_Z + \mathcal{H}_Q + \mathcal{H}_{HF} + \mathcal{H}_{CS} + \mathcal{H}_{D} + \mathcal{H}_J +\mathcal{H}_{other}  .
\label{NuclearHamiltonian}
\end{equation}
Here, $\mathcal{H}_Z$ and $\mathcal{H}_Q$ stand for the dominant Zeeman and quadrupolar interaction terms, respectively. The next four terms,  $\mathcal{H}_{HF}$, $\mathcal{H}_{CS}$, $ \mathcal{H}_{D} $, and $ \mathcal{H}_J$, are hyperfine interaction,  the chemical shift, dipole-dipole interaction, and $J$-coupling, respectively, and their relevance is material-specific. The last term $\mathcal{H}_{other}$ includes any potential time-independent interactions. The Zeeman term represents the direct coupling between the nuclear intrinsic magnetic moment $\gamma \hslash \mathbf{I}$ and the magnetic field $\mathbf{B}_0$ externally applied in the laboratory:
\begin{equation}
\mathcal{H}_Z = -\gamma \hslash \mathbf{I} \cdot \mathbf{B}_0, 
\label{ZeemanTerm}
\end{equation}
where $\gamma$ is the gyromagnetic ratio of the spin, while $\hslash \mathbf{I}$ is the spin operator of the nucleus.
The term $\mathcal{H}_Q$ represents the interaction between the electric quadrupole moment of the nucleus and the electric field gradient (EFG), generated by the surrounding 
electrons, where 
\begin{equation}
	H_Q = \frac{e Q}{2I(2I-1)} \cdot \mathbf{I} \mathbf{V}(\Theta) \cdot \mathbf{I},
\end{equation}
for the EFG tensor $\mathbf{V}(\Theta)$, given by 
\begin{equation}
	\bm{\mathbf{V}} = \begin{pmatrix}
		\partial_{xx} V & \partial_{xy} V & \partial_{xz} V\\
		\partial_{yx} V & \partial_{yy} V & \partial_{yz}  V\\
		\partial_{zx} V & \partial_{zy} V & \partial_{zz} V
	\end{pmatrix},
\end{equation}
for $\partial_{ij}V \equiv \frac{\partial^2 V}{\partial x_i \partial x_j}$. In the coordinate system of the principal axis of the EFG it reads:
\begin{equation}
\mathcal{H}_Q =
\frac{e^2 q Q}{4I(2I-1)}
\left( 3 I_Z^2 - I(I+1) + \frac{1}{2} \eta ( I_+^2 + I_-^2 ) \right)
\label{QuadrupoleTerm}
\end{equation}
where $I$ is the nuclear spin number, $e$ is the elementary charge, $eq=V_{ZZ}$ is the largest eigenvalue of the EFG tensor, $eQ$ is the electric quadrupole moment, and $\eta$ is the asymmetry parameter of the EFG. In strongly correlated materials, the next most important term is the hyperfine coupling,  which describes the interaction of the nuclear spin with the electronic spin that includes a dipole-dipole interaction and potential Fermi contact term, given by 
	\begin{equation}
		\label{eq:hyperfineFULL}
		\mathcal{H}_{HF} = \mathbf{S}\tilde{A}\mathbf{I},
	\end{equation}
where $\mathbf{S},\mathbf{I} $ is the electronic/nuclear spin operator, respectively, and $\tilde{A}$ is the hyperfine tensor \cite{Abragam}. The other terms and their secular approximations are described in  \ref{appx:hamiltonians}. 

In NMR/NQR, one probes these interactions by sending a pulse of radiation onto the system, which accounts for a perturbing term to be included in the full Hamiltonian:
\begin{equation}
\mathcal{H}_{1} (t) = \left( 2 \mathbf{B}_1 \cos (2 \pi \nu_P t - \varphi_P) \right) \cdot \mathbf{I}
\label{PulseTerm}
\end{equation}
where $2 \mathbf{B}_1$ is the magnetic component of the radiation pulse, $\nu_P$ and $\varphi_P$ are its frequency and phase,  respectively. This  $B_1$ radiation field is in a plane perpendicular to the externally applied magnetic field, $B_0$, that defines the Zeeman quantization axis.

Before the application of any pulses, the system is   in its thermal equilibrium state, which at room temperature can be approximated as $ \rho (0) = \exp (-\mathcal{H}_0/k_B T) / \mathcal{Z} = (1 - \mathcal{H}_0/k_B T) / \mathcal{Z} $. Computing the evolution of this state under the action of a pulse is equivalent to finding the corresponding evolution operator $U(t_P, 0)$, where $t_P$ is the time duration of the pulse. In the most general case, this operator cannot be computed directly, since the full Hamiltonian $\mathcal{H} = \mathcal{H}_0 + \mathcal{H}_1 (t)$ may depend on time. Here, $\mathcal{H}_0$ usually encompasses only terms of the full Hamiltonian particular to the problem, that our software supports. 

\subsection{Computing the spin dynamics}

The program has three main modes of evaluating the dynamics of the system: 
\begin{enumerate}[$i$]
	\item Direct diagonalization of the unitary evolution operator.
	
	 The method is intended for for low-dimensional spin systems and time-independent Hamiltonians, where pulses are modeled as instantaneous rotation operators.
	\item Average Hamiltonian Theory (AHT) up to 3${}^\text{rd}$ order in the Magnus expansion.

The AHT approach is appropriate for time-dependent Hamiltonains whenever the Zeeman interaction is dominant, ensuring that the Magnus expansion converges.
	\item Using QuTiP's solvers which support collapse operators and non-unitary evolution.
	
	This is most general and resource-demanding method that can handle time-dependent Hamiltonians. The QuTiP backend's efficiency can be improved if one compiles with the optional Cython and parallelization dependencies.
\end{enumerate}
Depending on the intended implementation,  a user can readily choose the optimal mode for the appropriate evaluation of the spin dynamics. We point out that a full simulation of any realistic material is impossible because of the huge dimension of the resulting Hilbert space ($2^N$, where $N$ is the number of interacting nuclei). Instead, spin dynamics are modeled by effective spin components that can be calculated in a highly reduced Hilbert space of just a handful of spins. \\


\noindent{\it   i.   Direct Diagonalization}

Even though this  high-precision, high-performance solver is designed for time-independent Hamiltonians,   it   supports the modeling of NMR pulses as instantaneous operators that represent a spin rotation.  The advantage of this approach is that the precision is independent of the time steps used. The floating point precision is the only limiting factor because the dynamics of the system are  governed by the time evolution operator, $U(t)$, for the initial state, $\ket{\psi(0)}$,   of the following form,
 \begin{equation}
	\psi(t) = U(t)\psi(0) = e^{-i \mathcal{H}t\hbar}\ket{\psi(0)},
\end{equation}
while in the density matrix formalism, one can utilize the von Neumann equation for the initial density matrix, $\rho(0)$,
\begin{equation}
	\rho(t) =  e^{-i \mathcal{H}t\hbar}\rho(0) e^{i \mathcal{H}t\hbar}.
\end{equation}
Such dynamics can then be simulated by executing the matrix diagonalization directly.   This method is a powerful and efficient tool for a fairly good approximation of  NMR dynamics. 

The direct diagonalization method can also handle RF pulses if they are considered as idealized spin rotation operators. Thus, a pulse of angle $\alpha$, applied along some direction is represented by 
\begin{equation}
    \mathcal{R(\alpha)} = \exp\{-i\alpha \hat{I}\},
\end{equation}
where $\hat{I}$ is the axis along which the spins are rotated. We note that one could model the time-dependent RF pulses within the exact diagonalization approach by dividing the Hamiltonian into small time-steps in which each Hamiltonian is treated as time-independent. However, if the time-discretization is too fine, the exact diagonalization approach will no longer have a performance improvement over average Hamiltonian theory. Therefore,  we recommend that the Magnus expansion be used to investigate the effect of finite pulses.

 The implementation of quantum computing protocols on electron-nuclear systems, readily attainable  by using the hyperfine coupling (Eq.\ \ref{eq:hyperfineFULL}) \cite{SuterNV20}, represents an example of  a functional  application of the Direct Diagonalization mode. Successful coherent control of such electron-nuclear systems achieved through a highly-detailed simulation of the system's dynamics can facilitate the realization of robust quantum gates.  
 Such a simulation,  in some measure,  is difficult because of the three orders of magnitude difference  between nuclear and electronic spin  gyromagnetic ratios. 
 That is, using numerical methods to simulate the dynamics would require the use of a time-step size congruent with the particular timescale of the system, namely  in the case when  the Zeeman interaction is  dominant, as is usually the case for  systems  placed in a strong magnetic field. The    Zeeman interaction of electron-nuclear system  is represented by,  
\begin{equation}
	\mathcal{H}_0	= - \omega_n ( I_{z}  \otimes  \mathbbm{1})-\omega_s(\mathbbm{1} \otimes S_z),
\end{equation}
 where $\omega_n, \omega_s$ is the nuclear and electronic Larmor frequencies, respectively. The  standard NMR procedure would be to solve the dynamics of the system in the rotating frame of the nucleus. However, since the electronic Larmor frequency is   three orders of magnitude greater than the nuclear one, passing to the rotating frame of the nucleus is to no avail. Employing the direct diagonalization mode, one can surpass these issues because the dynamics of the system can be evaluated at each step, independent of any other. Any digitization issues can easily be overcome by including more points in the time array.  The parallelization in Python can be used to enhance  the program's performance, if speed-up is necessary. 

On the other hand,  the direct diagonalization method can become time-consuming as the dimensions of the relevant Hilbert space increase. This problem can be somewhat overcome by the parallelization process in Python. Nevertheless, direct diagonalization is practically impossible for systems with large Hilbert spaces, and additionally modeling dissipation would be a daunting task. For such endeavors, we turn to alternative numerical methods.\\ \\
\noindent{\it  ii.  Average Hamiltonian Theory}\newline
In most NMR applications, the dominant term is the Zeeman interaction, $\mathcal{H}_0 = \mathcal{H}_Z$, known as the \textit{rotating frame} \cite{Blanes, Slichter}. The procedure  that   this mode follows consists of two steps: 
\begin{enumerate}[i.]
\item The problem is cast to the interaction frame, where the only relevant term of the Hamiltonian is the perturbing one:
\begin{equation}
\mathcal{H}(t) \rightarrow \mathcal{H}_{\mathcal{H}_0} (t) =
\exp(i \mathcal{H}_0 t/\hslash) \mathcal{H}_1(t)
\exp(-i \mathcal{H}_0 t/\hslash)
\label{IPHamiltonian}.
\end{equation}
\item The evolution operator is approximated by retaining the number of terms of the Magnus expansion depending on the application \cite{Blanes, Slichter}, the first few of which are computed through the following formulas:
\begin{equation}
\label{Magnus1stTerm}
\Omega_1(t_P, 0) =
\int_{0}^{t_P} dt_1 \widetilde{\mathcal{H}}(t_1)
\end{equation}
\begin{equation}
\label{Magnus2ndTerm}
\Omega_2(t_P, 0) =
\frac{1}{2}
\int_{0}^{t_P} dt_1
\int_{0}^{t_1} dt_2
[\widetilde{\mathcal{H}}(t_1), \widetilde{\mathcal{H}}(t_2)]
\end{equation}
\begin{multline}
	\label{Magnus3rdTerm}
	\Omega_3(t_P, 0) =
\frac{1}{6}
\int_{0}^{t_P} dt_1
\int_{0}^{t_1} dt_2
\int_{0}^{t_2} dt_3
 \Big(
[\widetilde{\mathcal{H}}(t_1),
[\widetilde{\mathcal{H}}(t_2), \widetilde{\mathcal{H}}(t_3)]]
+ \\
[\widetilde{\mathcal{H}}(t_3),
[\widetilde{\mathcal{H}}(t_2), \widetilde{\mathcal{H}}(t_1)]]\Big)
\end{multline}
where $\widetilde{\mathcal{H}} \equiv -i \mathcal{H}(t)/\hslash$, from which the evolution operator is readily computed up to $n^\text{th}$ order, as 
\begin{equation}
U(t_P, 0) = \exp \Big(\Omega_1(t_P, 0) + \Omega_2(t_P, 0) + \Omega_3(t_P, 0)) + \cdots + \Omega_n(t_P, 0)\Big) .
\label{EOApproximateSolution}
\end{equation}
\end{enumerate}
The Magnus expansion has been implemented up to the 3$^\text{rd}$ order in the program as it was deemed adequate for a fast converging Hamiltonian, but this can be easily expanded up to the n$^\text{th}$ order.\\

\noindent{\it iii.  Subroutined QuTiP Solvers}

To simulate open quantum systems, we have incorporated QuTiP \cite{Johansson2012} subroutines into the PULSEE solver modules. Developed as an open-source framework, efficient and highly-optimized numerical simulator, this package includes different equations, such as the Schr\"odinger's equation, Lindblad Master equation, Bloch-Redfield master equation, and a Stochastic Solver. The user can choose which one they deem most appropriate for the NMR simulation at hand, bearing in mind that some of these solvers are resource-intensive. 

Using QuTiP  circumvents the time-independent limitation of the direct diagonalization methods. Furthermore, it   improves on the Magnus expansion, which converges poorly in some interaction regimes. What is more, QuTiP allows PULSEE to take advantage of collapse operators for dissipation through the Lindblad Master equation to properly account for dephasing, instead of  an empirical decay function, such as the loss of magnetization via a function $\mathcal{M}(t, T_2)$ described below, used in other methods.

\subsection{Simulating NMR observables}
 
 The PULSEE package can be used  to generate and analyze the typical observable measured in magnetic resonance  experiments, such as the free induction decay signal (FID). In laboratories, the FID is the electrical signal  induced in a coil wound around the sample after the  electromagnetic RF pulse which generates the $B_1$ field is switched off. This signal is simply related to the component of the sample's magnetization along the axis of the coil, related to the FID itself \cite{Hore}. If the coil is oriented along $\mathbf{\hat{n}}$, then the FID signal will be given by:
 
\begin{equation}
S(t) = \mathrm{Tr} \left[
\rho (t)
\mathbf{\hat{n}} \cdot
\mathbf{I}
\mathcal{M}(t, T_2)
\right] \qquad t > t_P
\label{FIDSignal}
\end{equation}
where we   replaced the magnetization with the spin operator $\mathbf{I}$ of a single nucleus in the ensemble, since they are equal up to a scaling factor, and we have introduced an empirical functional form of the loss of magnetization, or the decay of signal, $\mathcal{M}(t, T_2)$, usually set to $\exp(-t/T_2)$. One can directly specify the form of the function, for example programming a stretched exponential, or pass in as many parameters (decoherence times $T_2$, stretching exponents $\beta$, etc) as necessary to mimic desired decays. In effect, our software generates a complex FID whose imaginary part represents the signal induced in an additional coil orthogonal to $\mathbf{\hat{n}}$, so that in the most common situation the FID reads, $S(t) = \mathrm{Tr} \left[ \rho (t) I_+\mathcal{M}(t, T_2) \right]$.

Once the  FID is acquired, one typically computes its Fourier transform to obtain what is called the NMR/NQR spectrum. This is the main outcome of the experiment and provides information about the interactions experienced by the system and the transitions that occurred in its evolution. This is  shown by the expansion of the FID in Fourier components:
\begin{equation}
S(t) =
\displaystyle\sum_{\varepsilon, \eta}
\mel{\varepsilon}{I_+}{\eta} \mel{\eta}{\rho(t_P)}{\varepsilon}
\exp (i \omega_{\varepsilon, \eta} t)
\label{ExpansionIx}
\end{equation}
where $\varepsilon$, $\eta$ run over the energy eigenvalues of the system, $\ket{\varepsilon}$, $\ket{\eta}$ are the corresponding eigenstates, and $\omega_{\varepsilon, \eta}$ is the frequency of transition between these two. This formula proves that the peaks of the NMR/NQR spectrum are located at the resonance frequencies $\omega_{\varepsilon, \eta}$, and that some of these frequencies may not show up in the spectrum if the associated transition has not occurred or if the detection setup is not oriented appropriately. In addition, comparison with the experimental spectrum may reveal the deviation of the actual system from the idealized, theoretical case, thus generating a basis for further study.

\section{Structure and usage of the software}
\label{SUSoft}

PULSEE is not simply a simulator of the time evolution of nuclear spin states, but it also reproduces   all the main features of NMR/NQR experiments. In this way, the program is a valuable tool in    experimental research, as the outcomes of a simulation are generated in a form that  can be directly compared with the results measured in a laboratory. 

Numerical simulations are prone to errors due to assumptions in approximations and in numerical absolute tolerance if pushed beyond their intended use. In order to ensure  full control over PULSEE and a   reliable reproduction of results, we have opted for a completely independent implementation, in addition  to an integration with an   already-exiting framework, such as QuTiP \cite{Johansson2012},  to fully grasp any  potential numerical artifacts. In particular, we noticed that the 2$^\text{nd}$ order Magnus expansion was insufficient in the interaction (Dirac) frame, but sufficient in the rotating reference frame (RRF), induced by $\mathcal{O}_{RRF}=h\nu I_z$. By going to a higher order in the Magnus expansion, the two pictures converged, demonstrating the necessity of being able to fully access and change the source-code of the program. Moreover, there is great benefit to incorporating PULSEE and QuTiP because of the relevant extra features already developed within this framework.

 Another source of error in NMR simulation software is the discrepancy between simulated and measured results arising from deviations from idealized/instantaneous pulses and the absence of noise normally encountered in experiments. PULSEE allows both the effect of the finite pulse and the noise on observables to be investigated. Specifically, we address the effects of the pulse duration  by evolving the system under the influence of the relevant Hamiltonian for the appropriate time that corresponds to the desired pulse. Such an evolution introduces noise, especially when handling more complicated interactions. Because instantaneous pulses are pertinent  for the QIS community, we have   developed a module  that allows for the simulation of idealized quantum circuits and gates (Sect.\ \ref{sec:quantum_circuits}). Furthermore, PULSEE can be deployed to simulate field inhomogeneities,  as well as distributions in different parameters, such as the quadrupolar coupling term, and  the Zeeman term, by averaging over multiple Hamiltonians, effectively simulating environmental noise. This is another functional feature which was included to assist in the design of optimal noise spectroscopy protocols \cite{McCoy1989, D.K.1998, Ferrand2015, Ajoy2019, Sung2021}.

\subsection{Download, dependencies and launching}
\label{Download}

The software can be downloaded from the following GitHub repository:  \url{https://github.com/vemiBGH/PULSEE}

PULSEE has been written entirely in Python 3.7. One must install PULSEE by navigating to the directory where the file setup.py is located, and by running
\begin{verbatim}
  $ pip install -e.
 \end{verbatim}
The program makes wide use of many of the standard Python modules (namely \texttt{numpy}, \texttt{scipy}, \texttt{pandas}, \texttt{matplotlib}) for its general purposes, \ as well as the Quantum toolkit in Python (\texttt{QuTiP}) \cite{Johansson2012}. We strongly recommend using the Anaconda distribution. Tests have been carried out using the \texttt{pytest} framework and the \texttt{hypothesis} module. The software includes a GUI which has been implemented with the tools provided by the Python library \texttt{kivy}. 
In addition, it is highly recommended that QuTiP's \texttt{parallel computation} module is used as it dramatically reduces the runtime, especially for the direct diagonalization method, by spawning processes and fully leveraging multiple processors on a given machine.

The two different GUIs are launched from the directory \texttt{src/pulsee} by entering the following command in the terminal
\begin{verbatim}
  $ python PULSEE_CMP_GUI.py
  $ python PULSEE_CHEM_GUI.py
\end{verbatim}
\noindent The use of the GUI is strongly discouraged and only suggested as a `quick-and-dirty' modeling technique. Otherwise, one  is strongly advised to use the functions defined in the module \texttt{Simulation} to write a custom simulation, as outlined in subsection \ref{BuildSimulation}. To give more freedom to the user and appeal to a wider audience more familiar to Mathematica, we have written Jupyter notebook demos that are easy to adapt to the system under investigation.

\subsection{Modules of the software}
\label{Modules}

The program consists of 6 modules. Below, the content and role of each module is briefly described:

\begin{enumerate}
\item \texttt{Operators}

This module, together with \texttt{Many\_Body}, is to be considered a 
toolkit for the simulation of generic quantum systems. It contains the definition of Python classes and functions related to the basic mathematical objects which enter the treatment of a quantum system. \texttt{Operators} simulates of a single spin system, while \texttt{Many\_Body} extends it to systems made up of several spins.

\item \texttt{Many\_Body}

This module contains the definitions of the functions \texttt{tensor\_product} and \texttt{partial} \texttt{\_trace} which allows  a single particle Hilbert space to pass to a many particle space, and vice-versa.

\item \texttt{Nuclear\_Spin}

This module is where the classes representing the spin of an atomic nucleus or a system of nuclei are defined.

\item \texttt{Hamiltonians}

This file includes the definition of the relevant terms of the Hamiltonian of a nuclear spin system in a typical NMR/NQR experiment: namely, these are the Zeeman interaction, quadrupolar interaction, full $J$-coupling between nuclei using the $J$ tensor, isotropic chemical shift in the secular approximation, dipolar for homonuclear spins, dipolar for heteronuclear spins, hyperfine interaction in the secular approximation, any interaction that can represented with a tensor between two spins, $J$-coupling in the secular approximation, and interaction with an RF pulse of radiation. Finally, the program allows the input of a square matrix as a numpy array that represented any predefined Hamiltonian

\item \texttt{Simulation}

This is the module the user should refer to in order to implement a custom simulation. The functions defined here allow the user to set up the nuclear system, evolve it under the action of a sequence of pulses, generate the FID signal, and compute the NMR spectrum from it.

\item\texttt{Quantum\_computing}

This module contains the implementations of fundamental components of quantum 
circuits, including several quantum gates, and \texttt{Qubit} objects, acted upon by gates. In principle, the user may construct and manipulate elementary quantum circuits, and extract relevant information, such the final density matrix of the composite qubit state.

\item \texttt{NMR\_NQR\_GUI}

This is the graphical user interface (GUI) of the program. There are two versions of the GUI depending on the application. The condensed matter physics (CMP) \texttt{PULSEE\_CMP\_GUI} deals only with a single-spin system that is governed by the Zeeman \& Quadrupolar interactions. The second \texttt{PULSEE\_CHEM\_GUI} extends the single-spin system to include weaker couplings in the secular approximations, mostly useful for applications in chemistry, by considering a generalized secondary spin.    Although the GUI provides a simple and intuitive way to perform a simulation, it has limited features with respect to a custom simulation code in order to reproduce complicated experiments involving long multi pulse sequences.
\end{enumerate}

\subsection{Building up a simulation}
\label{BuildSimulation}

The starting point of any simulation is the set up of the system under study, which is done by calling the function \texttt{nuclear\_system\_setup}:
\begin{verbatim}
nuclear_system_setup(spin_par, quad_par=None, zeem_par=None, \
					 j_matrix=None, cs_param=None, D1_param=None, \
					 D2_param=None, hf_param=None, h_tensor_inter=None, \
					 j_sec_param=None, h_userDef=None, \
					 initial_state='canonical', temperature=1e-4)
\end{verbatim}
This function returns three objects representing the spin system, the unperturbed Hamiltonian, and the initial state, respectively.

The next step is to evolve the state of the system under the action of a pulse of radiation, a task carried out by the function \texttt{evolve}:
\begin{verbatim}
evolve(spin, h_unperturbed, dm_initial, \
       mode=None, pulse_time=0, \
       picture='RRF', RRF_par={'nu_RRF': 0,
                               'theta_RRF': 0,
                               'phi_RRF': 0}):
\end{verbatim}
The function not only allows the user to specify the features of the pulse applied to the system, but also the reference frame where the evolution is computed.

Once  the evolved state is obtained as the outcome of \texttt{evolve}, one can generate the FID signal associated with this state by calling the function \texttt{FID\_signal}:
\begin{verbatim}
FID_signal(spin, h_unperturbed, dm, acquisition_time, T2=100, \
           theta=0, phi=0, reference_frequency=0)
\end{verbatim}
The arguments of this function allow the user to set the time window of acquisition of the FID, the decoherence time $T_2$, the orientation, and the frequency of rotation of the detecting coils.

Eventually, one computes the NMR/NQR spectrum from the FID signal by passing the FID through  the function \texttt{fourier\_transform\_signal}: 
\begin{verbatim}
fourier_transform_signal(times, signal, frequency_start, \
                         frequency_stop, opposite_frequency=False)
\end{verbatim}

The module \texttt{Simulation} also includes  the functions for plotting the density matrix of the evolved state as well as the FID signal and NMR/NQR spectrum.

\section{Examples of execution}
\label{ExEx}

In this section we illustrate some noteworthy simulations performed with PULSEE. 
In addition to   being valid examples of the execution of the code, these simulations have been chosen because they clearly demonstrate the precision of NMR and NQR in the control of   nuclear spin degrees of freedom, which reflects their accuracy in the determination of  unknown nuclear interactions in a sample under study.

\subsection{Selective transitions between quadrupolar states by means of properly polarized pulses in NQR experiments}

The structure of the energy spectrum of quadrupolar nuclei allows for the selective excitation of its states by applying a pulse of radiation with the proper polarization.

\begin{figure}[t]
\centering
    \includegraphics[scale=0.4]{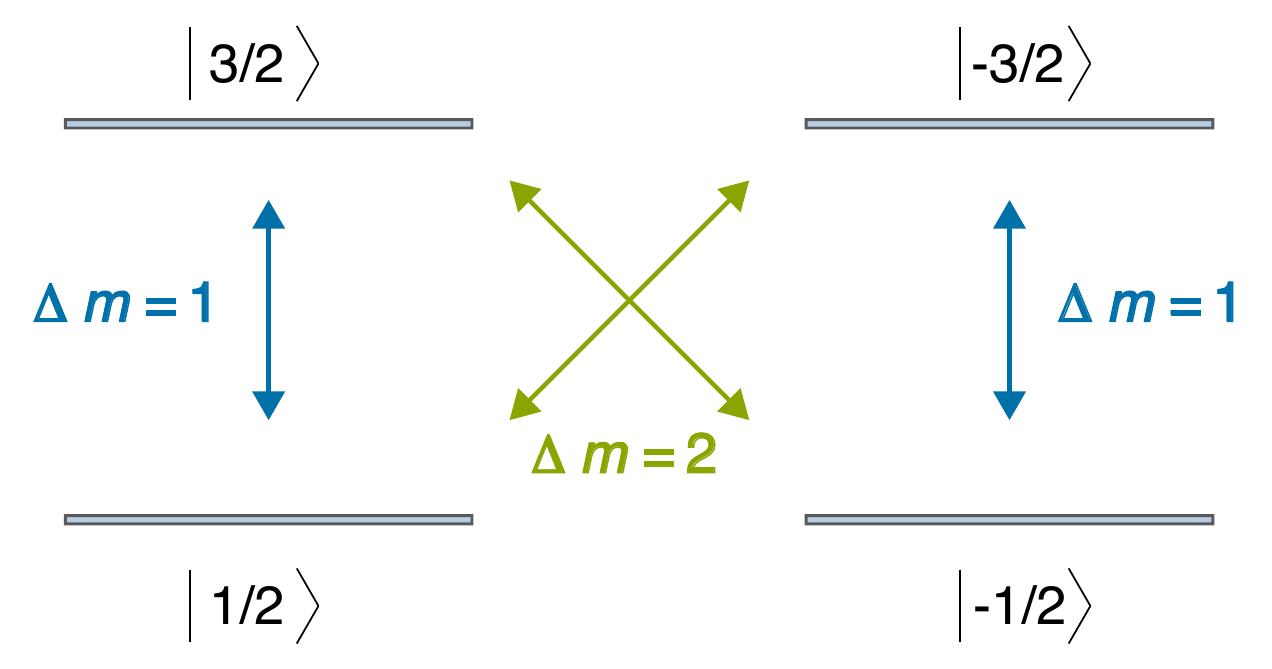}
    \caption{\small{Scheme of the energy spectrum and the available transitions of a quadrupolar nucleus of spin 3/2. The transitions labeled with $\lvert \Delta m \rvert = 1$ involve the exchange of a single photon, while those labeled $\lvert \Delta m \rvert = 2$ involve two photons.}}
    \label{Quadrupole32Transitions}
\end{figure}

A first notable example is represented by the pure NQR of a spin 3/2 nuclei  whose energy levels and available transitions are depicted in \mbox{Figure \ref{Quadrupole32Transitions}}. This system may undergo two single photon transitions at the same frequency, namely $\ket{1/2} \leftrightarrow \ket{3/2}$ and $\ket{-1/2} \leftrightarrow \ket{-3/2}$. This is  in contrast with a pure NMR experiment  where all the transitions are characterized by the same variation of the magnetic quantum number $\Delta m$.  These two transitions imply an opposite change in the angular momentum of the system, so that each of them can occur only under the exchange of a photon with circular polarization (c.p.) $\sigma^+$ and $\sigma^-$, respectively. Therefore, when one irradiates the system by a linearly polarized (l.p.) resonant pulse, both transitions will be induced.  In contrast, by choosing the proper polarization of the pulse one is able to select only one of the two. The potential of circularly and, in general, elliptically polarized RF pulses in NQR has been widely explored \cite{Weber1960, LEE2001355, MILLER2001228}.

\begin{figure}[t]
\centering
    \includegraphics[scale=0.50]{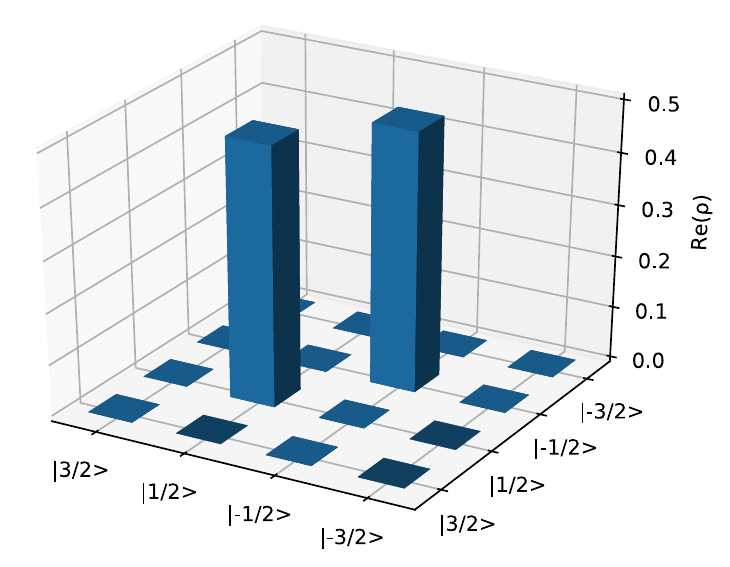}
    \caption{\small{Real part of the density matrix representing the initial state of the spin 3/2 nucleus in the simulation of a pure NQR experiment. The system is prepared in a classical distribution where the states at the ground level are equally populated.}}
    \label{PureNQR32InitialState}
\end{figure}

\begin{figure}[t]
\centering
    \includegraphics[width=0.98\textwidth]{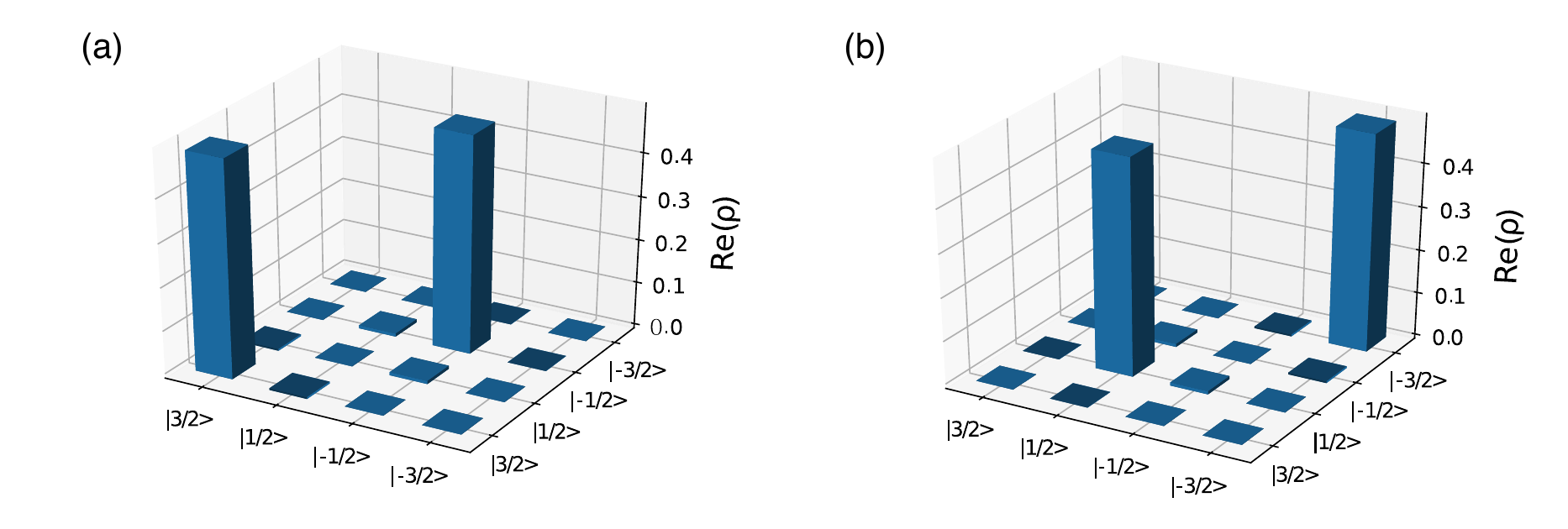}
    \caption{\small{Real part of the simulated density matrix representing the evolved state of the spin 3/2 quadrupolar nucleus after: (a) a $\pi$ pulse with polarization $\sigma^+$; (b) a $\pi$ pulse with polarization $\sigma^-$.}}
    \label{PureNQR32EvolvedStates}
\end{figure}

These theoretical expectations are correctly reproduced by our software. We simulated the pure NQR of a spin 3/2 \textsuperscript{35}Cl nuclei in a potassium chlorate crystal (KClO\textsubscript{3}), whose gyromagnetic ratio is $\gamma/2\pi = 4.17$ MHz/T and whose  quadrupolar resonance frequency is $\nu_Q = 28.1$ MHz \cite{Das}. We prepared the system in the initial state depicted in Figure \ref{PureNQR32InitialState}. Then, we performed two distinct simulations evolving the system under the action of a $\pi$ pulse with polarization $\sigma^+$ or $\sigma^-$ respectively (in a classical picture such pulses rotate the initial nuclear magnetization by 180$^{\circ}$,  clockwise and anticlockwise,  respectively).  The results   obtained are shown in Figure \ref{PureNQR32EvolvedStates}.
We note that the  $\pi$ pulse is defined such that its amplitude, $B_1$, and time duration, $t_P$, satisfy the equation known as a central-transition selective pulse
\begin{equation}
\gamma \alpha B_1 t_P = \pi
\label{PiPulseEquation}
\end{equation}
where $\alpha = \sqrt{I(I+1)-m(m+1)}$  is a factor depending on the transition being induced, and differs between the central and satellite peaks, and  $\gamma$ is the gyromagnetic ratio of the nucleus. For the pulsing to be successful, the strength of the applied pulse $\gamma B_1$ must be smaller than the quadrupolar frequency $\omega_Q$. If the applied pulse is not perturbative relative to the quadrupolar energies, then the levels will mix instead of only rotating. 

The evolved density matrices clearly show  that a pulse with circular polarization $\sigma^{+}$ ($\sigma^{-}$) couples only with the transition between states $\ket{1/2} \leftrightarrow \ket{3/2}$ ($\ket{-1/2} \leftrightarrow \ket{-3/2}$) by acting selectively on the two relevant  energy eigenstates to induce a full inversion of their respective  populations 
 in such a way that the total angular momentum is conserved. 

\begin{figure}[t]
\centering
    \includegraphics[scale=0.45]{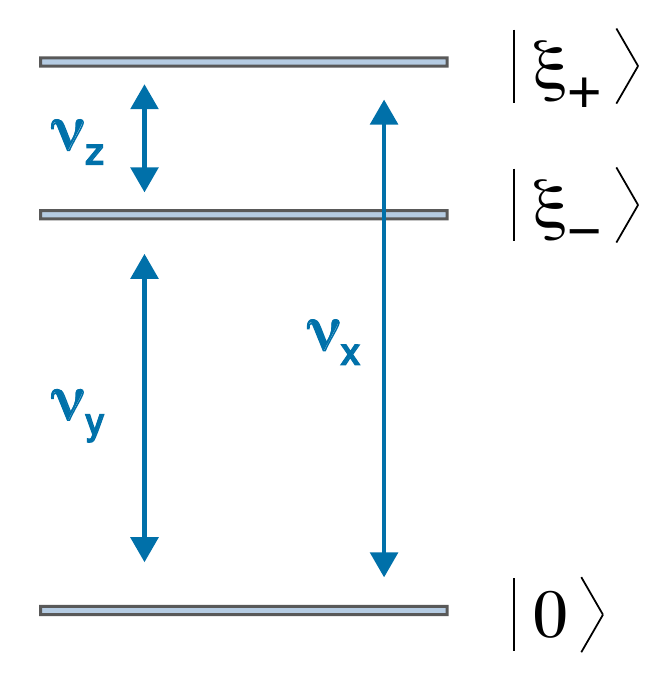}
    \caption{\small{Scheme of the energy spectrum and the available transitions for a quadrupolar nucleus of spin 1 with an asymmetric EFG. The subscripts of the transition frequencies $\nu_{x/y/z}$ refer to the direction of linear polarization of the pulse required to induce each transition.}}
    \label{Quadrupole1AsymTransitions}
\end{figure}

Another experiment where the pulse can be set up to selectively induce transitions  is the NQR of a spin 1 nucleus in the presence of an asymmetric EFG \cite{Abragam}. Due to the non-vanishing asymmetry parameter, the energy eigenstates of this system are no longer the spin eigenstates, but they read:
\begin{equation}
\ket{0}  \qquad \& \qquad
\ket{\xi_\pm} = \left( \ket{1} \pm \ket{-1} \right)/\sqrt{2}
\label{I=1AsymmetricEnergyLevels}
\end{equation}
The energies of these states and the frequencies of transitions between them are displayed in Figure \ref{Quadrupole1AsymTransitions}.

What is peculiar with this system is that in order for each of the three transitions to occur, the pulse must have a distinct linear polarization. Indeed, developing the calculations one finds that
\begin{equation}
I_x =
\begin{blockarray}{cccc}
\ket{\xi_+} & \ket{0} & \ket{\xi_-} &  \\
\begin{block}{(ccc)c}
0 & 1 & 0 & \ket{\xi_+} \\
1 & 0 & 0 & \ket{0} \\
0 & 0 & 0 & \ket{\xi_-} \\
\end{block}
\end{blockarray}
\quad
I_y =
\left(
\begin{array}{ccc}
0 & 0 & 0 \\
0 & 0 & 1 \\
0 & 1 & 0 \\
\end{array}
\right)
\quad
I_z =
\left(
\begin{array}{ccc}
0 & 0 & 1 \\
0 & 0 & 0 \\
1 & 0 & 0 \\
\end{array}
\right)
\label{PureNQRAsymSelectivePulses}
\end{equation}
from which it is easy to prove that an $\mathbf{\hat{x}}$-, $\mathbf{\hat{y}}$- or $\mathbf{\hat{z}}$-polarized pulse will only affect the transition $\ket{\xi_+} \leftrightarrow \ket{0}$, $\ket{\xi_-} \leftrightarrow \ket{0}$, or $\ket{\xi_+} \leftrightarrow \ket{\xi_-}$, respectively. This behavior can be assessed by observing the spectrum generated by each of these 
pulses,    recalling  \mbox{Eq. \eqref{ExpansionIx}}. Once the proper orientation of the detection coils is set, one is able to visualize if a certain transition has occurred  depending on the whether the term $\mel{\eta}{\rho(t_P)}{\varepsilon}$ vanishes or not.

\begin{figure}[t]
\centering
    \includegraphics[scale=0.5]{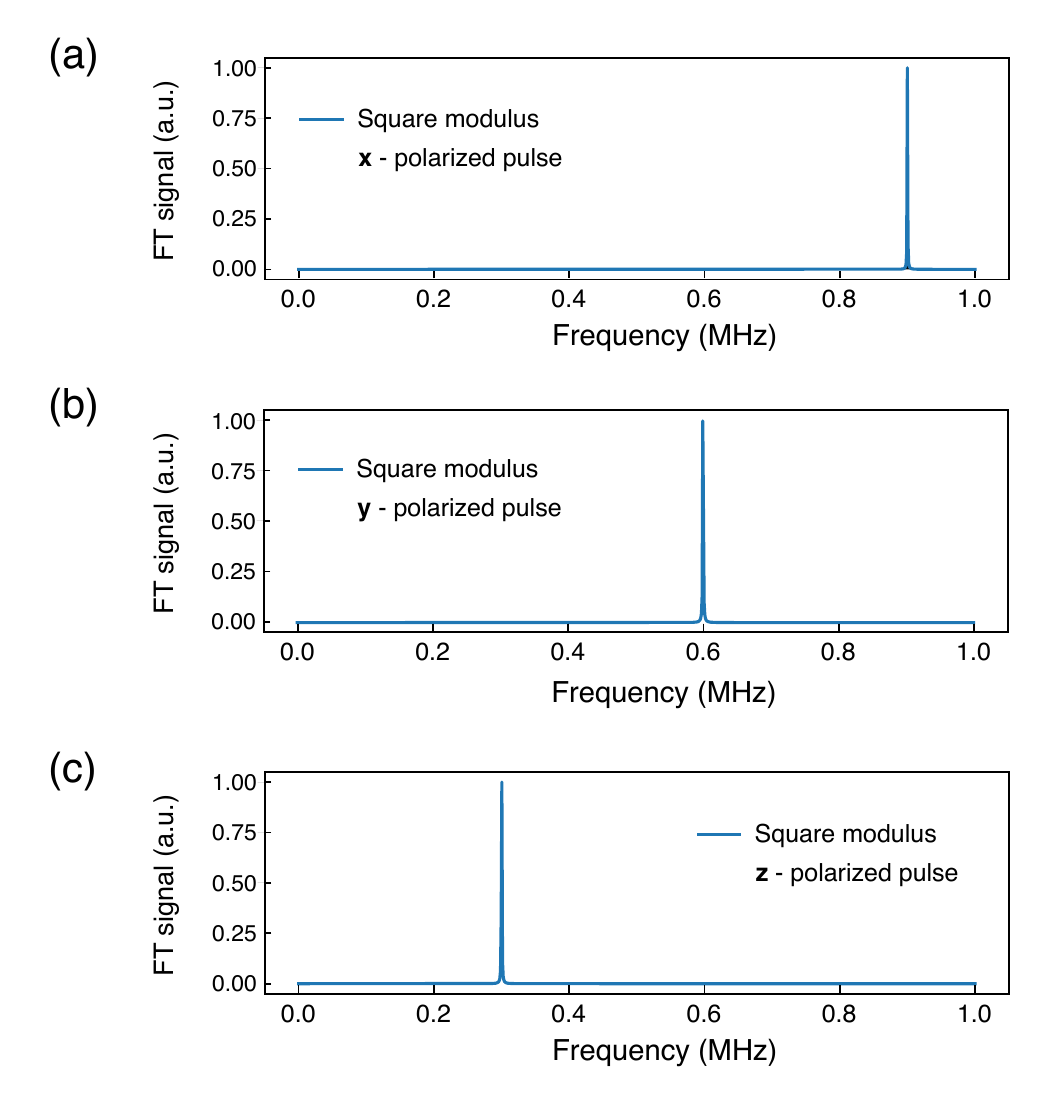}
    \caption{\small{Spectra resulting from three distinct simulations of the NQR of a spin 1 nucleus in an asymmetric EFG, where different pulses have been applied with polarization along $\mathbf{\hat{x}}$ (a), $\mathbf{\hat{y}}$ (b), and $\mathbf{\hat{z}}$ (c), respectively.}}
    \label{PureNQR32AsymSpectra}
\end{figure}

These results have been simulated in a fictitious spin 1 nucleus with $e^2qQ/h = 1\, \rm{MHz}$ and asymmetry $\eta = 0.6$, for which the transition frequencies are $\nu_x \equiv \nu_{\xi_+ \leftrightarrow 0} = 0.9\, \rm{MHz}$, $\nu_y \equiv \nu_{\xi_- \leftrightarrow 0} = 0.6\, \rm{MHz}$,  and $\nu_z \equiv \nu_{\xi_+ \leftrightarrow \xi_-} = 0.3 \, \rm{MHz}$. The resulting NQR spectra are displayed in Figure \ref{PureNQR32AsymSpectra}.

\subsection{Generation of quantum coherences in a spin 3/2 quadrupolar nucleus}

Let us consider the spin 3/2 \textsuperscript{35}Cl nucleus in the same KClO\textsubscript{3} crystal introduced above. In the previous example, we showed how to induce a full inversion of the populations of two of its energy eigenstates, say $\ket{m=1/2}$ and $\ket{m+1=3/2}$, by means of a $\sigma^+$ c.p. $\pi$ pulse of radiation. 
 In general,
when the angle on the right-hand side of \mbox{Eq. \eqref{PiPulseEquation}} is set to a value different from $n\pi$, where $n$ is an integer, the final density matrix exhibits non-zero off-diagonal elements, meaning that the evolved state includes a quantum superposition of $\ket{m}$ and $\ket{m+1}$. 
 Such superposition states can be deployed to probe nature of tensor multipolar orders \cite{RongOrbit19, PhysRevLett.127.237201}. 
In NMR, such elements are typically called ``single quantum coherences'', where ``single'' specifies the fact that $\Delta m = 1$.

\begin{figure}[t]
\centering
    \includegraphics[scale=0.332]{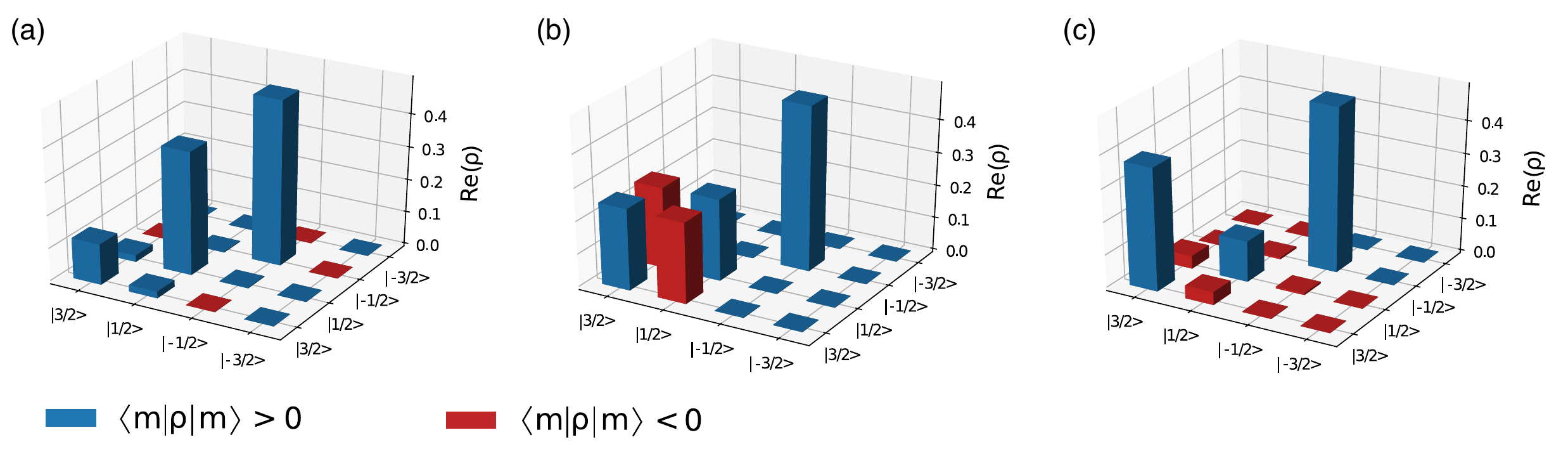}
    \caption{\small{Results of the simulation of the NQR of \textsuperscript{35}Cl nuclei in a KClO\textsubscript{3} crystal where a pulse resonant with the $\ket{1/2} \leftrightarrow \ket{3/2}$ transition is applied. The three histograms show the real part of the density matrix of the system evolved after   (a)   $\pi/3$ pulse, (b)   $\pi/2$ pulse, and  (c)   $2\pi/3$ pulse. As the angle of rotation approaches $\pi$, the populations of the states involved in the transition undergo a continuous exchange, and at the same time non-zero,  off-diagonal elements emerge between them, meaning that the two states are in a quantum superposition.}}
    \label{NQRQuantumCoherences}
\end{figure}

In Figure \ref{NQRQuantumCoherences},  we display the results of three simulated experiments where a pulse resonant with the $\ket{1/2} \leftrightarrow \ket{3/2}$ transition is applied and the angle in \mbox{Eq.\ \eqref{PiPulseEquation}}  is set to the values $\pi/3$, $\pi/2$,  and $2\pi/3$, respectively.  These simulations demonstrate that it is possible to fine-tune the amplitudes of two states linked by a single-photon transition through the careful manipulation of the parameters of the pulse.

\subsection{Preparation of an ensemble of spin 3/2 nuclei in a pseudopure state by means of NQR}
\label{PseudoPure}

NMR and NQR are naturally suited for  the implementation of simple quantum information processors, as they are an efficient and high-precision method for   manipulating  nuclear spins \cite{Cory, Murali, Kampermann, NMRQIS_Book, Oliveira2012}. Nonetheless, in typical NMR/NQR experiments the system under study is a macroscopic sample made up of a huge number of nuclei, which makes it impossible to prepare it in a pure state as would be required by an ordinary quantum computation protocol.
This problem has been addressed by following a different strategy \cite{Cory97}: by means of a properly designed pulse sequence, the ensemble of nuclear spins can be prepared in a pseudopure state, \ie a state $\rho = a \mathbbm{1} + b\ketbra{\psi}{\psi}$ which differs from a pure state by a term proportional to the identity.  A state of this kind is called pseudopure because, under evolution, it behaves like a full-fledged pure state. This property makes it the ideal starting point for any NMR/NQR quantum computation protocol. Indeed, much effort has been made in realizing quantum logic gates \cite{Jones1998, Price1999, Dorai, Sinha01, Ramanathan2004, Guelec2005, Teles2012, Tan2012, Teles2015, Wolfowicz2016, Min2022}. 

\begin{figure}[t]
\centering
    \includegraphics[width=0.65\textwidth]{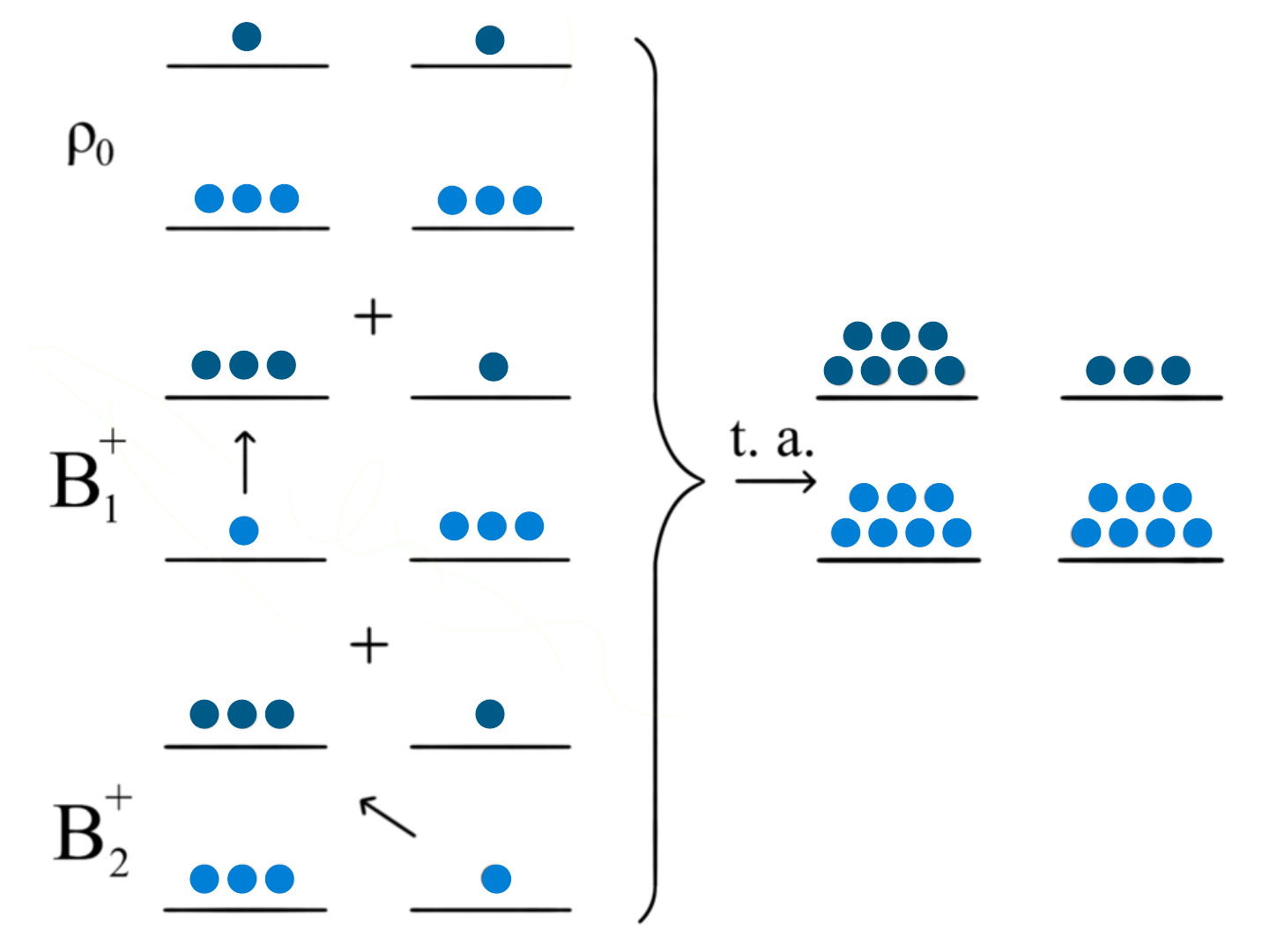}
    \caption{\small{Population diagrams of the states to be combined through temporal averaging in order to realise the $\ket{11}=\ket{-3/2}$ pseudopure state in a spin 3/2 quadrupolar nucleus.}}
    \label{PseudoPurePopAdjustDiagram}
\end{figure}

In what follows, we describe a simulation of the NQR protocol aimed at realizing a 2-qubit pseudopure state in the ensemble of spin 3/2 quadrupolar \textsuperscript{35}Cl nuclei of a KClO\textsubscript{3} crystal \cite{Possa}, whose energy levels and available transitions have already been  shown in Figure \ref{Quadrupole32Transitions}. The states of the 2-qubit computational basis correspond to the spin ones as follows:
\begin{equation}
\begin{array}{ccc}
\ket{00} & \equiv & \ket{3/2} \\
\ket{01} & \equiv & \ket{1/2} \\
\ket{10} & \equiv & \ket{-1/2} \\
\ket{11} & \equiv & \ket{-3/2}. \\
\end{array}
\label{TwoQubitSpinStatesCorrespondence}
\end{equation}
Here, we remark that the simulated protocol is not aimed at implementing the pseudopure state in the physical system itself.  Such a state is obtained as the average of the results of three distinct experiments, as depicted in \mbox{Figure \ref{PseudoPurePopAdjustDiagram}}, following a common practice employed in NMR/NQR called temporal averaging \cite{Knill98}. In each of the three experiments, the system is handled in a distinct way:
\begin{enumerate}
\item In the first, the system is left in its original thermal equilibrium state.

\item In the second, the system is irradiated by a c.p. pulse with resonant frequency $\nu_Q = (E_{\pm 3/2} - E_{\pm 1/2})/h$,  inducing one of the single photon transitions ($\lvert \Delta m = 1 \rvert$). The time duration of the pulse is set to a value such that the populations of the states linked by the transition are exchanged. 

\item In the third, a c.p. pulse at half the resonance frequency is applied, yielding one of the two-photons transitions ($\lvert \Delta m = 2 \rvert$). Again, the time duration of the pulse accounts for the exchange of the populations of the states linked by the transition.
\end{enumerate}

\begin{figure}[t]
\centering
    \includegraphics[width=0.8\textwidth]{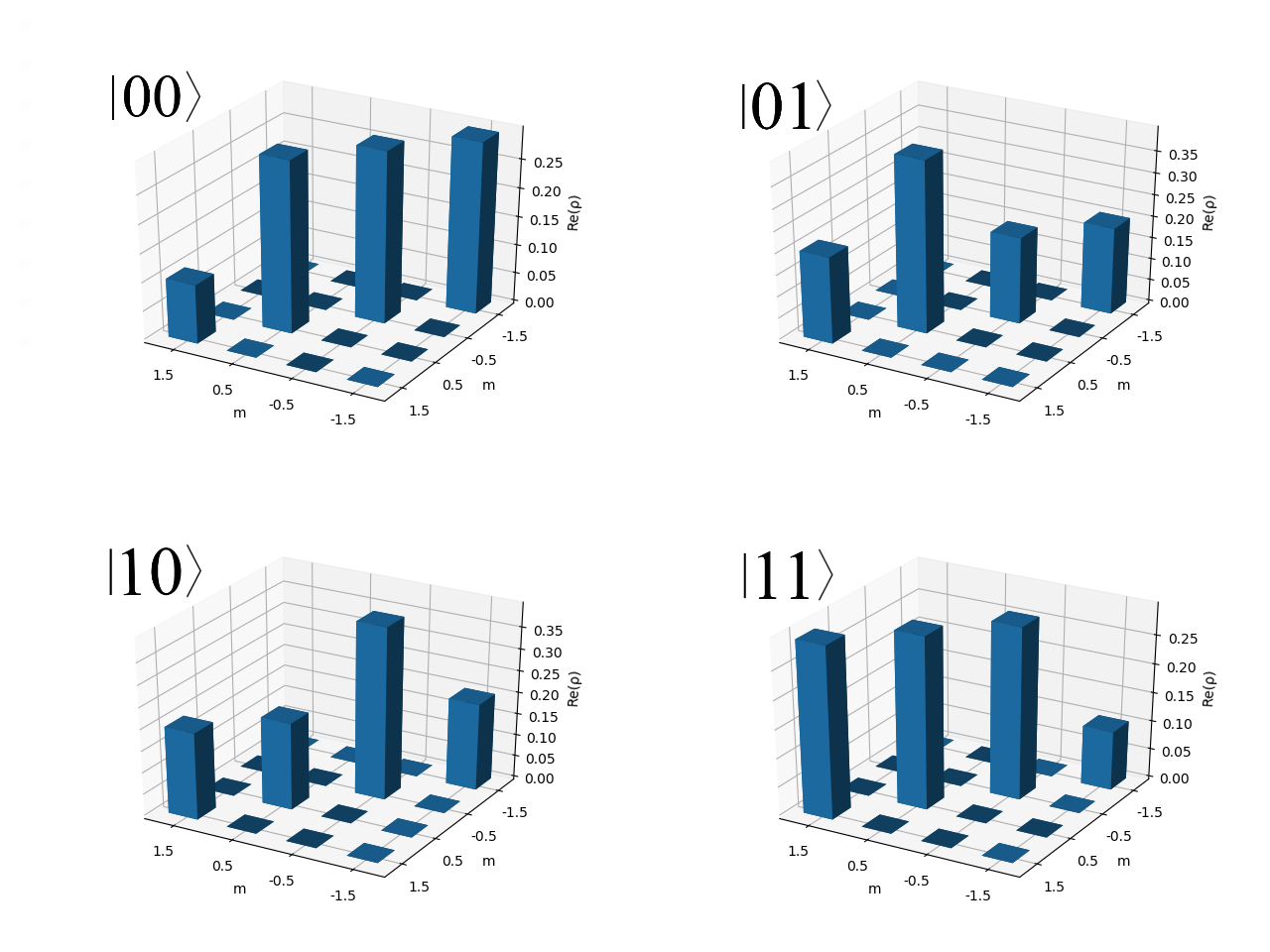}
    \caption{\small{Outcomes of the protocol for the realization of pseudopure states in an ensemble of spin 3/2 quadrupolar nuclei, as simulated by our software. The histograms display the real part of the density matrices representing the four pseudopure states of the computational basis of 2 qubits.}}
    \label{PseudoPure32NQRResults}
\end{figure}

If
 the polarization of the pulses applied in   steps 2  and 3  are appropriately chosen,    the average of the density matrices resulting from the three experiments will have the properties of a pseudopure state belonging to the computational basis in \mbox{Eq.\ \eqref{TwoQubitSpinStatesCorrespondence}}. The outcomes of the simulation, illustrating the real part of the density matrices representing the four pseudopure states of the computational basis of 2 qubits,  are shown in \mbox{Figure \ref{PseudoPure32NQRResults}}.

\subsection{NQR and NMR implementation of a CNOT gate on a couple of qubits}
\label{CNOT}

\begin{figure}[t]
\centering
    \includegraphics[scale=0.26]{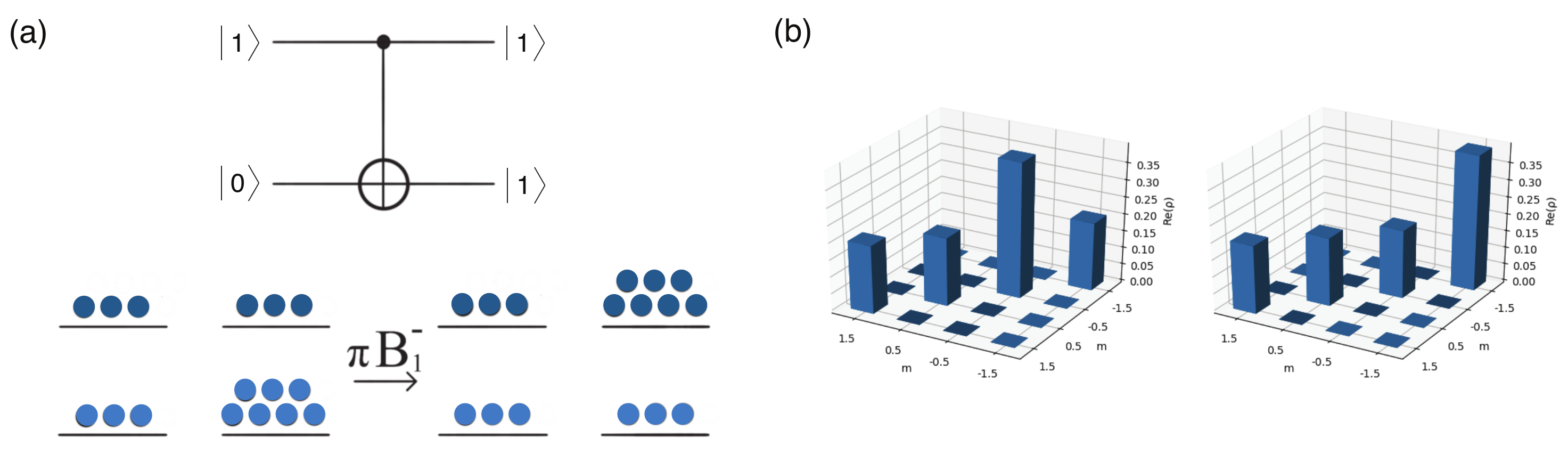}
    \caption{\small{{\bf (a)} Symbolic notation of a CNOT gate operating on the state $\ket{10}$ (on top) and the action of the pulse which carries out the equivalent operation on the NQR version of the 2-qubit system (at the bottom). {\bf (b)} Input (on the left) and output (on the right) states of the simulated NQR CNOT\textsubscript{1} gate when the initial state is $\ket{10}$.}}
    \label{NQRCNOTGateDiagram}
\end{figure}

Implementing a CNOT gate in the system we have already discussed in subsection \ref{PseudoPure} is a straightforward task. 
That is, in a 2-qubit system with $\ket{0}$ and $\ket{1}$ as the only allowed input values for both qubits, the CNOT gate flips the second (target) qubit from $\ket{0}$ to $\ket{1}$ if and only if the first (control) qubit is in the $\ket{1}$ initial state.
Indeed, one can easily check that the action performed by a CNOT\textsubscript{1} gate on the 2-qubit system is equivalent to that of a pulse which yields an exchange of the populations of the states $\ket{-1/2}$ and $\ket{-3/2}$, as illustrated in \mbox{Figure \ref{NQRCNOTGateDiagram}a}.  The effect of this gate as simulated by our software and is depicted in \mbox{Figure \ref{NQRCNOTGateDiagram}b}.

\begin{figure}[h]
\centering
    \includegraphics[width=0.9\textwidth]{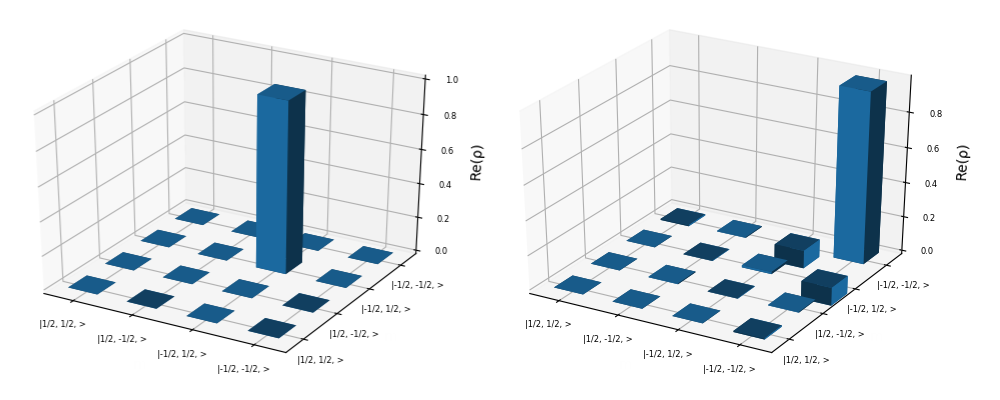}
    \caption{\small{Input (on the left) and output (on the right) of the simulated NMR CNOT\textsubscript{1} gate when the initial state is set to the ideal pure state $\ket{10}$. The output state presents a slight discrepancy with respect to the expected $\ket{11}$ state, which is thought to be a consequence of the discrete approximations taken in the simulation.}}
    \label{NMRCNOTGate10}
\end{figure}

It is possible to implement an analogous operation by means of NMR as well, but in a different nuclear system. As explained in \cite{NMRQIS_Book}, this time the 2 qubits are encoded in 2 distinct spin 1/2 nuclei (following the convention $\ket{0} \equiv \ket{1/2}$, $\ket{1} \equiv \ket{-1/2}$) and, in order for them to work as a control-target qubit couple,  they must interact with each other. Thus, we assume that they are linked by the typical $J$-coupling, whose contribution to the Hamiltonian is:
\begin{equation}
\mathcal{H}_{J} = h J I_{z}^{(1)} I_{z}^{(2)}
\label{JCoupling}
\end{equation}
where $J$ is the coupling constant and $I_{z}^{(i)}$ is the $z$ component of the spin of the $i$-th nucleus.
The experimental protocol for the implementation of an NMR CNOT gate employs both selective rotations of each spin as well as the free evolution of the whole system under the action of $J$-coupling, according to the sequence:
\begin{equation}
\text{CNOT}_{1} = \left( - \frac{\pi}{2} \right)^{\mathbf{I}_1}_{z}
\left( \frac{\pi}{2} \right)^{\mathbf{I}_2}_{z}
\left( - \frac{\pi}{2} \right)^{\mathbf{I}_2}_{x}
U \left( \frac{1}{2J} \right)
\left( - \frac{\pi}{2} \right)^{\mathbf{I}_2}_{y}
\label{CNOTGate}
\end{equation}
Here, factors of the type $\left( \alpha \right)^{\mathbf{I}_i}_{x/y/z}$ represent pulses resonant with the $i$-th spin which make it rotate   an angle $\alpha$ around the axis specified in the subscript. $U \left( 1/2J \right)$, on the other hand, stands for the free evolution of the system for a time duration of $1/2J$.
We point out that in order to be able to perform selective rotations of one of the two spins, the nuclei's gyromagnetic ratios must be appreciably different, leading to well separated gyromagnetic frequencies $\nu_0^{(i)} = - (\gamma^{(i)}/2\pi) B_0$.

We have carried out a simulation of this protocol starting from ideal pure input states. The outcomes match closely our expectations, \ie the  initial  ket      is flipped since the control (second) qubit was in the $\ket{1}$ initial state, as is shown in \mbox{Figure \ref{NMRCNOTGate10}}.

\subsection{NMR probe of quantum correlations and tensor orders}  
\label{HyperFSim}
As a local probe,   NMR is well suited for the  study of the microscopic electronic spin structure in the vicinity of  the nuclear spin site through the hyperfine interaction. While directly measuring quantum correlations between electronic spins is difficult, complex hyperfine interactions can imprint signatures of electronic correlations on the nuclear spin states. The resulting many-body nuclear spin correlations can then be probed using the method of multiple quantum NMR \cite{Feldman2008, Gerasev2018, Gaerttner2018}. The main challenge with this effort is that nuclear spin states are not pure states precluding the direct application of standard quantum protocols, which can be addressed by using pseudo-pure states (sec.\ \ref{PseudoPure}). 
PULSEE can be instrumental in designing the optimal multiple quantum NMR sequence to permit the study of quantum correlations. 

   Many of the theoretically identified complex  quantum phases of   materials  are characterized by tensor orders (e.g. ferro-octupolar order) \cite{PhysRevLett.127.237201, Pourovskiie21} that possess zero local susceptibility, and for that reason, are evasive to standard experimental probes. However, the tensor nature of the hyperfine interactions can reveal  the intricate structure of quantum orders. In this section, we illustrate ways in which PULSEE is deployed   to put forward a novel NMR method, inspired by QIS, that allows for the  engineering of  pulse sequences that can effectively probe   electronic correlations and tensor orders    through   the hyperfine interaction.

In order to explore the capabilities of PULSEE, we   give a simple yet powerful illustration of two approaches to modeling the hyperfine interaction. In the first, we  considering a spin-1/2 nucleus coupled to an electronic bath  directly  
 via a hyperfine interaction ($\mathcal{H}_{\rm hf} =  \mathbf{S}\tilde{A}\mathbf{I}$).  In  the second, we  examine two spin-1/2 nuclei interacting via an effective hyperfine field, $\tilde{A}$, mediated via electrons  ($\mathcal{H}_{\rm hf} =  \mathbf{I_{1}}\tilde{A}\mathbf{I_{2}}$).  
The system will be modeled as two interacting spin-1/2 particles, governed by the Hamiltonian,  
\begin{equation}
	\label{eq:twospinHhyperfine}
	 \mathcal{H} = - \omega_n ( I^{(1)}_{z}  \otimes  \mathbbm{1}) +  
	\begin{cases}
	 -\omega_s(\mathbbm{1} \otimes S_z)  +  \mathbf{S}\tilde{A}\mathbf{I^{(1)}}, & \text{i}   \\\\
	 -\omega_n( \mathbbm{1} \otimes I^{(2)}_{z})  +  \mathbf{I^{(1)}}\tilde{A}\mathbf{I^{(2)}}, & \text{ii} 
	 \end{cases}
\end{equation}
 where  $I, S$ correspond to the nuclear/electronic spin operator, respectively. The electronic Larmor precession frequency  is much greater than the nuclear precession, $\omega_s \approx 2000 \omega_n$, and  the Zeeman terms dominate over the hyperfine coupling. 

The two forms of the hyperfine interactions are written for different applications. For instance, 
the Hamiltonian defined in \mbox{Eq. \ref{eq:twospinHhyperfine}i} may be useful in organic materials that  exhibit very rich phase diagrams induced by strong correlations \cite{PhysRevX.12.011016}. 
The concepts introduced in the study of open quantum systems \cite{Ticozzi2017} can be exploited to discern the nature of complex phases arising as a result  of strong correlations. 
That is,  a target system is identified as nuclear spins (e.g. $^{13}$C) coupled to the electronic bath via the hyperfine interaction, to an uncontrollable bath
 as correlated electron spins,   and to an engineered auxiliary system  as nuclear spins interacting via the dipole-dipole interactions. The target and auxiliary systems    share an entangled state, reflecting the nature of the electronic correlations we seek to identify. By simulating the form of the expected experimental results, one may deploy PULSEE to devise effective pulse sequences to probe the quantum orders in such correlated phases.   
Furthermore,  \mbox{Eq. \ref{eq:twospinHhyperfine}i} can serve as a starting point for  quantum control studies \cite{Ajoy2018, QMetrology20, ChenCurr17, PhysRevA.105.022428, PhysRevX.12.011016}. 
  
On the other hand, the Hamiltonian defined in \mbox{Eq. \ref{eq:twospinHhyperfine}ii} may be useful in studies of mean-field electronic correlations and  in the development of the probes of tensor order
\cite{LaflammeNNhf11}. 

\begin{figure}[t]
	\centering
	\includegraphics[width=.96\textwidth]{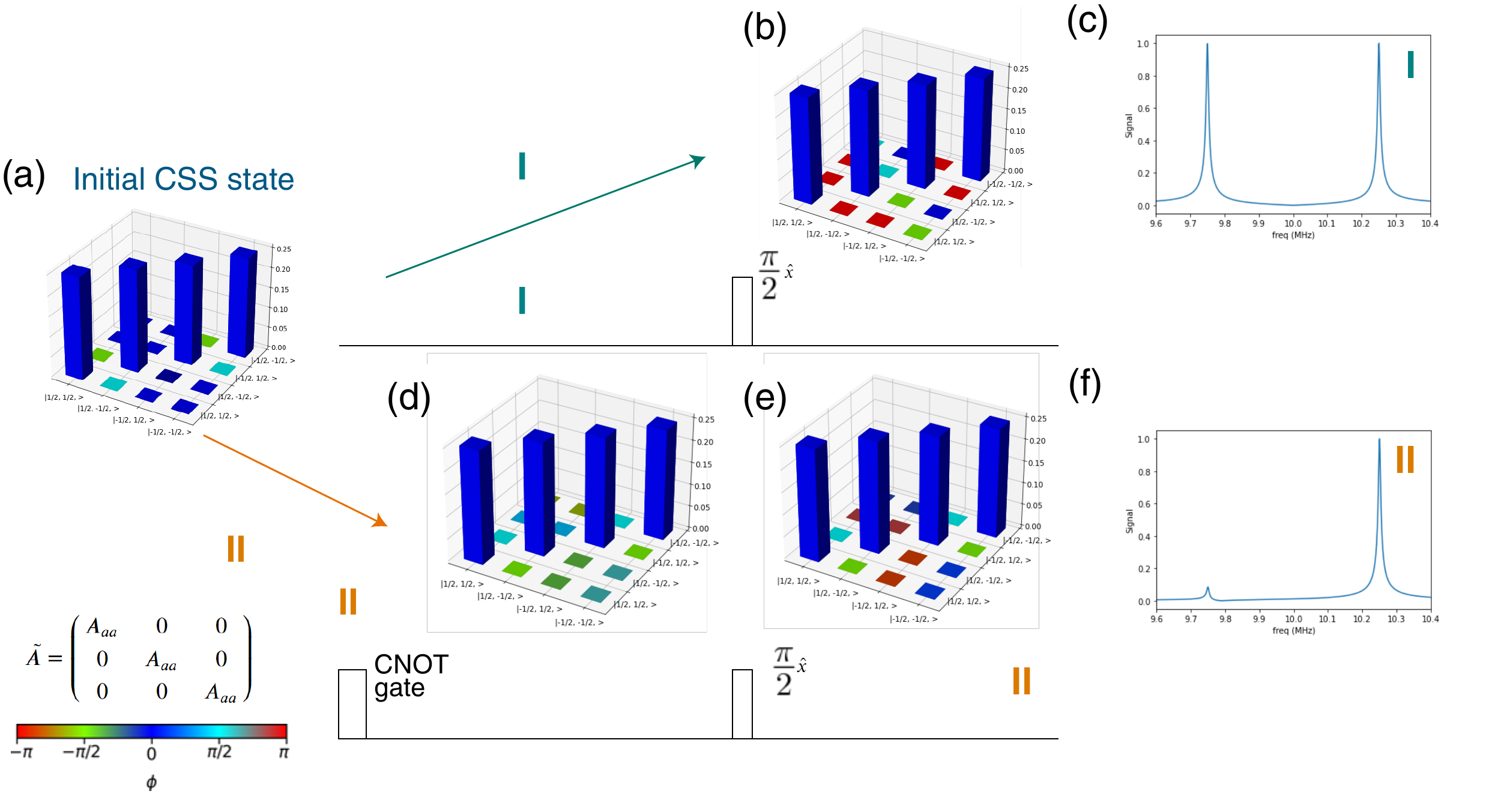}
	\caption{\small{Dynamics of the {\bf (a)} coherent spin state (CSS) (Eq.\ \ref{eq:intialCoherent}})  in the high temperature limit. 
	 The system is evolved under the Hamiltonian (Eq.\ \ref{eq:twospinHhyperfine} i), and a {\bf (b)} $\pi/2$ pulse along $I_x$ is applied to the nuclear spin, {\bf (c)}  observing the corresponding NMR spectrum  of the FID. The bottom row is a second experiment ({\bf II}), where a {\bf (d)}   CNOT gate (Eq.\ \ref{CNOTGate}) is applied {\bf (e)} before the $\pi/2$ pulse, {\bf (f)} after which the FID is observed. 
This shows that the CNOT gate can mimic the effect of the anisotropic hyperfine tensor on the CSS.}
	\label{CohDiagCNOT}
\end{figure}

 Here, we demonstrate the utility  of PULSEE in devising an efficient protocol to probe the nature of tensor order, \ie anisotropy of the hyperfine tensor. 
 We consider two spin-1/2 spins coupled  
via a hyperfine interaction of the form, 
\begin{equation}
	\label{eq:hyperfineTensorAac}
	\tilde{A} = 
	\begin{pmatrix}
		A_{aa} & 0 & A_{ac} \\ 
		0 & A_{aa} & 0 \\
		A_{ac} & 0 & A_{aa} 
	\end{pmatrix} 
\end{equation} 
where   $\tilde{A}$ is the second-rank hyperfine tensor representation for the  antiferromagnetic phase (AFM) with symmetry plane $y=0$ \cite{Koutroulakis10}. The diagonal terms ($A_{aa}$) of $\tilde{A}$ dominate, giving the principal axes. The system will be modeled as two interacting spin-1/2 particles, governed by the Hamiltonian \mbox{Eq. \ref{eq:twospinHhyperfine}i}.
Working in the Zeeman-dominant regime, we investigate the evolution of the coherent spin state (CSS), as we have identified  these as the most sensitive to anisotropy of the hyperfine tensor. Tuned to the nuclear spins, we can only probe the system by sending pulses to the nucleus. In the high-temperature limit, the thermal state of our system is given by
\begin{equation}
	\label{eq:intialThermal}
	\rho_\text{thermal state} \approx \alpha \mathbbm{1} - \epsilon_s S_z - \epsilon_n I_z,
\end{equation}
where $\epsilon_s = \frac{\hslash\omega_s}{k_BT}, \epsilon_n = \frac{\hslash\omega_n}{k_BT} \sim 10^{-6}  \ll 1$ are the polarization factors at room temperature $T$, with two different deviation density matrices for the uncorrelated electronic and nuclear spins,  $\alpha$ is a constant that depends on the temperature and Hamiltonian of the system, and where $\hslash/ k_B$ is the Planck/Boltzmann constant.  We work in   units of $\hslash = 1$. To obtain the CSS for the nucleus, one can transform the thermal state into a state of the form
\begin{equation}
	\label{eq:intialCoherent}
	\rho_\text{CSS nucleus} \approx\alpha  \mathbbm{1} - \epsilon_s \mu- \epsilon_n \sigma,
\end{equation}
where $\mu$ is the polarized state of the electron, given by $\mu = \ket{1}\bra{1}$, whenever the electrons are in a magnetically ordered state \cite{Koutroulakis10}, and $\sigma$ is the deviation matrix of the nuclear spin's CSS. The CSS saturates the Heisenberg uncertainty relation \cite{Radcliffe1971} and   resembles a semiclassical spin.   It  is of the form 
\begin{equation}
	\label{eq:theoreticalNCSC}
		\ket{\zeta(\theta,\varphi)} = \sum_{m=-j}^{j}  \binom{2j}{j+m}^{1/2}\cos^{j+m}\frac{\theta}{2}\sin^{j-m}\frac{\theta}{2}e^{i(j-m)\varphi}\ket{j,m},
\end{equation}
where $j$ is the nuclear spin number, which in our case is $j = 1/2$, $\theta, \varphi$ are angles in the Bloch sphere,  and $\ket{j,m}$ are the eigenstates of the $I_z$ operator \cite{Radcliffe1971}. In principle, these angle give the rotation of the $I_z$ operator, and the CSS are eigenstates of the rotated $I_z$ operator, namely
\begin{equation}
	( \vec{R}_{\theta\varphi}I_z \vec{R}_{\theta\varphi}^\dagger) \ket{\zeta(\theta,\varphi)} = j \ket{\zeta(\theta,\varphi)},
\end{equation}
for the rotation operator, $\vec{R}_{\theta\varphi} = e^{i\theta(I_x\sin\varphi-I_y\cos\varphi)}$. The angles $\theta = \pi/2$ and  $\varphi = \pi/2$ are chosen for the particular case when there is no squeezing and the squeezing parameter is unity \cite{AuccaiseEstrada2013, Auccaise2015}; thus the deviation matrix is $\sigma = \ket{\zeta(\pi/2,\pi/2)}\bra{\zeta(\pi/2,\pi/2)}$. These nuclear spin coherent pseudopure states (NSCS) have been experimentally prepared using the adapted strongly modulated pulse \cite{Auccaise2015}.  
\begin{figure}[t]
	\centering
	\includegraphics[width=.96\textwidth]{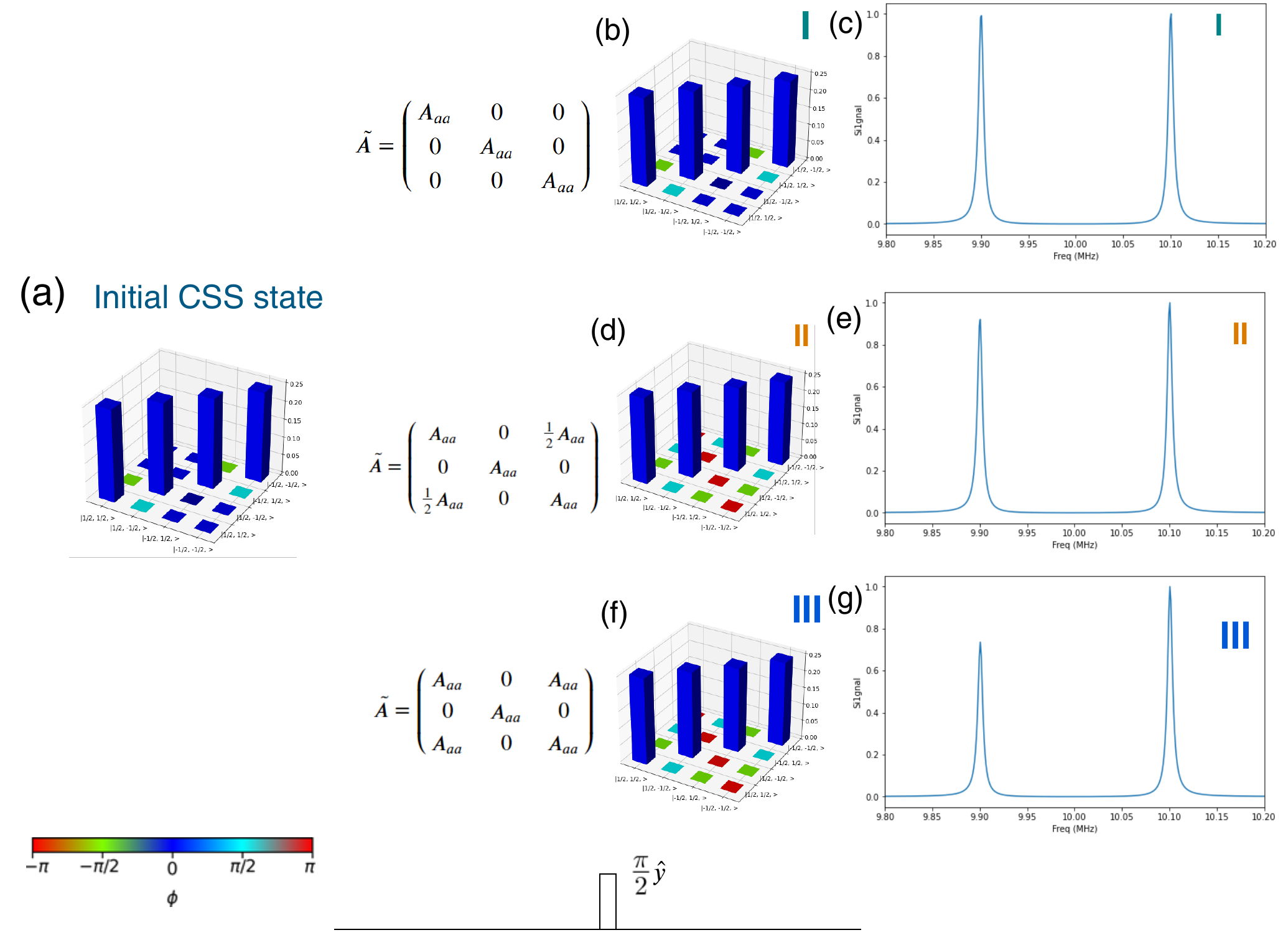}
	\caption{\small{Dynamics of the {\bf (a)} coherent spin state (CSS) (Eq.\ \ref{eq:intialCoherent}}) in the high temperature limit. The system is evolved under the Hamiltonian (Eq.\ \ref{eq:twospinHhyperfine} i), a {\bf (b)} $\pi/2$ pulse along $I_x$ is applied to the nuclear spin, {\bf (c)}  observing the corresponding NMR spectrum  of the FID. Three simulations are given for the three different forms of the hyperfine tensor, I fully diagonal, II off-diagonal term $A_{ac} = \frac{1}{10} A_{aa}$,   and III $ A_{ac} = \frac{1}{5} A_{aa}$. This CSS is sensitive to the anisotropy of the hyperfine tensor, acting as an effective probe of tensor orders. }
	\label{CoherentOffDiag}
\end{figure}
 We perform a typical FID experiment simulation to obtain the NMR spectrum by evolving the initial state under the hyperfine Hamiltonian in \mbox{Eq.\ \ref{eq:twospinHhyperfine}i} using the direct diagonalization method, applying a $\pi/2$ pulse to the nuclear spin, and observing the FID. 
We have assumed that  $B_0  = 10\, \rm{T}$  and $T_2 = 50 \,\mu \rm{s}$  and a 20 times longer acquisition time.  The hyperfine coupling is much weaker than the Zeeman term (on the order of few percent of the nuclear Zeeman term)  and it creates a peak splitting proportional to $A_{aa}$ (Fig.\ \ref{CoherentOffDiag}c \& Fig.\ \ref{CohDiagCNOT}c). 

 We consider three different hyperfine tensors of the form depicted in Eq.\ \ref{eq:hyperfineTensorAac}, with $A_{ac} = 0, A_{ac} = \frac{1}{2} A_{aa}$,  and $ A_{ac} = A_{aa}$. The form of the spectra and the evolved density matrices are given in \mbox{Fig.\ \ref{CoherentOffDiag}.} 
 Using PULSEE, we have explored the  sensitivity of various nuclear spin states to the form of the hyperfine tensor. 
 We found  that the particular  CSS (Eq.\ \ref{eq:intialCoherent}) is sensitive to the anisotropy of the hyperfine tensor. That is, the relative height of one of the peaks in the splitting changes as a function of the strength of the off-diagonal term $A_{ac}$. What is promising about this method is the fairly straight forward way to implement it experimentally. Once the correct CSS is prepared for the nucleus, the system is perturbed by a simple $\pi/2$ pulse along the appropriate axis, in this case $I_y$.  Our method is similar to the spin squeezing techniques in NMR pseudo-pure states \cite{Sinha2003}.

Working only with a diagonal hyperfine tensor, we show that a CNOT gate implementation (Eq.\ \ref{CNOTGate}) mimics the effects of the hypefine tensor (Fig.\ \ref{CohDiagCNOT}). In essence, the CNOT gate introduces entanglement,  where the first nuclear site is the ``control qubit'' and the second nuclear site is the ``target qubit.''

The combination of these two experiments gives us valuable information about the hyperfine tensor by studying the simple NMR spectrum.  Firstly, we see that the central line of the Zeeman spectrum is split, where the splitting is given by the parameter $A_{aa}$ of the hyperfine tensor. Furthermore, the application of the CNOT gate (Eq.\ \ref{CNOTGate}, where the last two $I_z$ pulses can be ignored, and $U = U(1/2A_{aa}$)) suppresses one of the peaks, Fig.\ \ref{CohDiagCNOT}f, as expected \cite{Teles2012, Teles2015}.
 Thorough investigation of the spin dynamics evolution after the application of the CNOT gate, allows us to establish the methodology for full hyperfine tensor determination.

Thus,  measurements on CSS states  serve as   control experiments to sense the anisotropic nature of the hyperfine interaction. 
In other words, by performing rather manageable experiments, one may determine the nature of the hyperfine interaction, that is, the presence of off-diagonal terms, without the need of full field rotation spectroscopy \cite{Lu17}. Even though this simple experiment is only tuned to the nucleus, one may envision different ways to couple to the electronic spin \cite{Ajoy2018, Liu2019}, and then use PULSEE to investigate the dynamics of the spin and the   observables. 


In summary, our software allows for the simulation  of complex spin evolution, 
which  may then be used to design  the appropriate pulse sequences
  enabling reverse engineering of the relevant Hamiltonians  of  tensor orders.

\subsection{Building quantum circuits module: correlated density matrices}
\label{sec:quantum_circuits}

\begin{figure}[t]
	\centering
	\includegraphics[width=.96\textwidth]{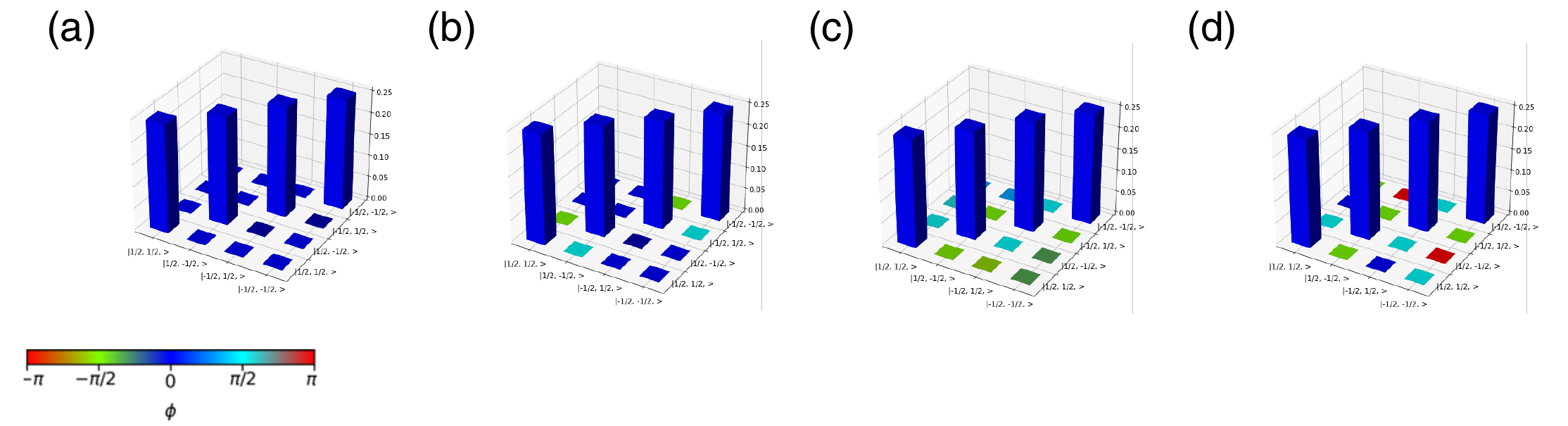}
	\caption{\small{Effects of the pulse artifacts on  the preparation of the nuclear spin coherent states (NSCS, Eq.\ \ref{eq:intialCoherent}) of a combined electronic-nuclear spin system, evolved under the second order average Hamiltonian theory mode. 
	   {\bf (a)} Density matrix of initial thermal state (Eq.\ \ref{eq:intialThermal}) is shown for comparison, where $\phi$ is the phase.  {\bf (b)}   Theoretical NSCS generated by instantaneous perfect pulses from Eq.\ \ref{eq:theoreticalNCSC}). NSCS prepared by applying a $\pi/2$ pulse along $I_x$, evolved under the {\bf (c)} Zeeman and  {\bf (d)} both Zeeman and hyperfine Hamiltonians (Eq.\ \ref{eq:twospinHhyperfine} $A_{ac}=0$). By simulating finite NMR pulses, the evolved density matrix deviates from the theoretical one, even in the simple Zeeman case without  any noise. Nevertheless, the gate fidelities (Eq.\ \ref{eq:fidelity}) of (c, d) are nearly unity. }}
	\label{fig:NonIdealCoh}
\end{figure}

The software supports designing quantum circuits via \texttt{QubitState} objects in the \texttt{Quantum\_computing} module, and tracking the dynamics of a density matrix as it evolves in the circuit. 
 Besides being useful for quantum circuit analysis, this module  is instrumental in investigating the effects of  experimental artifacts, such as pulse imperfections.   The effects of finite pulses applied in the lab cannot be equated with the those of instantaneous perfect gates. The artifacts of `imperfect' pulses  need to be considered when performing complex NMR pulse sequences. In order to evaluate the errors associated with finite pulses, one may consider the gate fidelity defined by \cite{Fortunato2002}  
\begin{equation}
	\label{eq:fidelity}
	F = \frac{\Tr (\rho_\text{th}\cdot\rho_\text{ex})}{\Tr(\rho_\text{th}\cdot\rho_\text{th}^\dagger)\Tr(\rho_\text{ex}\cdot\rho_\text{ex}^\dagger)},
\end{equation}
where $\rho_\text{th}$ is the theoretical density matrix, and $\rho_\text{ex}$ is the density matrix  obtained experimentally through quantum tomography \cite{Gaikwad2018}. Using PULSEE, one may test finite pulses, determine the level of additional terms in the density matrix, and determine  different pulse sequences and their fidelity in order to achieve  the most adequate pulse train for the desired state evolution. 

 In \mbox{Fig. \ref{fig:NonIdealCoh}} we illustrate the effect of the pulse artifacts on  preparation of the nuclear spin coherent states (NSCS) using the average Hamiltonian theory method.     
 The coherent spin state for a spin-1/2 particle whenever we use the angles $\theta = \pi/2$ and  $\varphi = \pi/2$ is $\ket{\zeta(\pi/2,\pi/2)} = \ket{+y}$, or the ground eigenstate of the $I_y$ operator. 
 In an NMR experiment,  this $I_y$ state is obtained from a thermal state following  the application a $\pi/2$ pulse along $I_x$.  However, this assumes that the Hamiltonian which governs the system is a simple Zeeman one, and the $\pi/2$ pulse is perfect. We examine the effect of non-ideal $\pi/2$  pulse  encountered when hyperfine interaction is present. Specifically,    we simulate the effects of a non-ideal $\pi/2$ pulse by evolving the initial thermal state under two different Hamiltonians (Zeeman and hyperfine) and assume that no other noise in present in the system. Although  the simulation does not include noise, we find that  the density matrices  differ when the full evolution of the pulse is considered under the different Hamiltonians, as depicted in \mbox{Fig. \ref{fig:NonIdealCoh}}.  However, we learned that their fidelities do not notably differ from unity. 
 These results demonstrate that in certain experiments one should examine the density matrices and not just consider fidelities to simulate proper  time evolution of the spins.
\begin{figure}[t]
	\centering
	\includegraphics[scale=0.33]{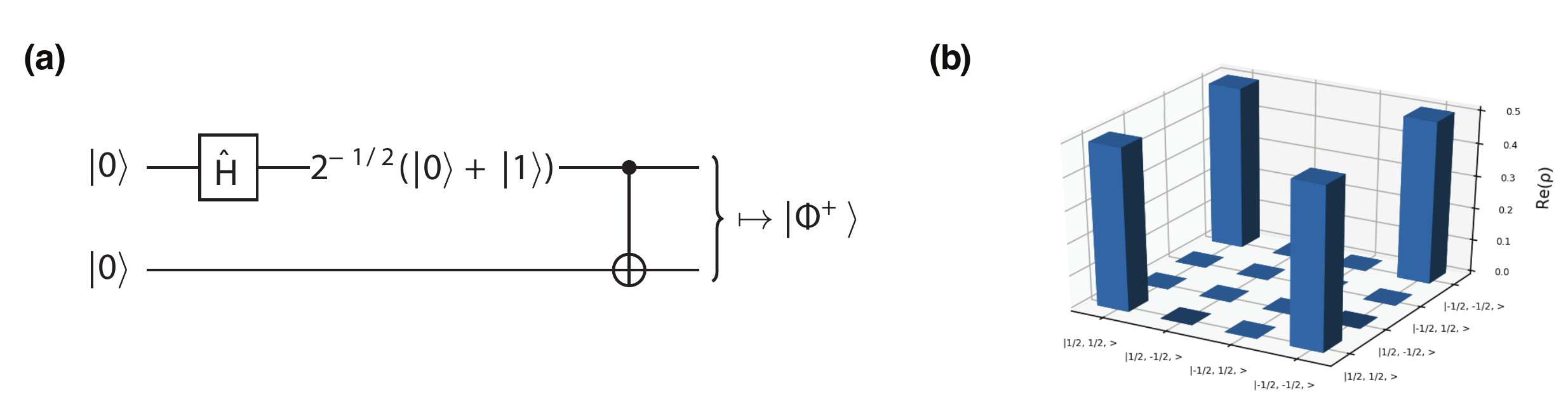}
	\caption{\small{{\bf (a)} Quantum circuit diagram of the application of a 
			Hadamard and CNOT gate to produce a Bell state from the computational 
			basis state $\ket{00}$.  {\bf (b)} Density matrix of correlated Bell state $\ket{\Phi_+} 
	 		= 2^{-1/2}(\ket{00} + \ket{11})$}.}
	\label{HadamardCNOTCircuit}
\end{figure}

 Theoretically predicted states can be modeled using the quantum computing module  as a benchmark with experimentally prepared density matrices. As an example of the quantum circuit builder, consider constructing a two-qubit, ``maximally-entangled'' Bell state, produced by applying a Hadamard gate to one qubit, which creates a superposition, and then subsequently applying a CNOT-gate, which entangles the two qubits by creating a control and a target qubit. The gates' matrix representations are 
\begin{equation}
	\hat{\text{H}} \equiv \frac{1}{\sqrt{2}}
	\begin{pmatrix}
		1 & 1 \\ 1 & -1 
	\end{pmatrix}   
	\quad \& \quad
	\text{CNOT}\equiv
	\begin{pmatrix}
		1 & 0 & 0 & 0 \\ 0 & 1 & 0 & 0 \\ 0 & 0 & 0 & 1 \\ 0 & 0 & 1 & 0 
	\end{pmatrix},
\end{equation}
in the computational basis. Taking the initial state as $\ket{00}$, one obtains
\begin{equation}
	\text{CNOT} [(\hat{\text{H}}\otimes \mathbbm{1}) \ket{00} ]
	= \text{CNOT} \left[\frac{1}{\sqrt{2}}(\ket{0} + \ket{1}) 
	\otimes \ket{0}]\right]
	= \frac{1}{\sqrt{2}} (\ket{00}+ \ket{11}),
\end{equation}
which is precisely the Bell basis state $ \ket{\Phi^+} $. The circuit is depicted in \mbox{Fig.\ \ref{HadamardCNOTCircuit}{\bf (a)}}, and the  density matrix  produced in 
\mbox{Fig.\ \ref{HadamardCNOTCircuit}{\bf (b)}}. 

Taking the control qubit as $A$, one may check that 
\begin{equation}
	\Tr(\rho_{\ket{\Phi^+}}^A) = 1/2 < 1, 
\end{equation}
confirming that this is indeed a correlated state \cite{Baaquie2013}.

\section{Conclusions} \label{Conc}

PULSEE is an open-source software for the simulation of nuclear magnetic resonance experiments  on complex materials. The main purpose of this program is to provide a numerical tool for the development of new methods of investigation of emergent properties in complex materials inspired by the NMR/NQR protocols established in the context of  quantum information processing \cite{NikolovAPS21}.

The software follows the principles of wide accessibility and intuitive utilization, being available for download from a public GitHub repository \cite{PULSEE} and providing a GUI, Jupyter notebooks,  as well as   complete and detailed documentation.

The examples of execution illustrate the features of the software, which include the ability to simulate both the evolution of spin states and the corresponding experimental observables, and also highlight the possibilities to manipulate nuclear spin states through NMR/NQR.  
 PULSEE enables simulations of the evolution of  a  single-spin under various interactions in solids. The investigation of  the  deviation of simulated results from  experimental  results on actual materials   through the subsequent inclusion of different interaction terms in the Hamiltonian, opens   up an opportunity to gain valuable insight into the microscopic nature of correlations in quantum materials.  
In that sense, our software might find its relevance in the design of highly sensitive protocols for the study of emergent quantum properties of materials.

\section{CRediT author statement}

\noindent{\bf Davide Candoli}: Investigation, Software Programming, Development and Validation, Formal analysis, Visualization, Data curation, Writing-Original draft preparation. 
{\bf Ilija Nikolov}:  Software Programming, Development and Validation (NMR probe of Quantum correlations and quantum gates), Data curation, Writing-Original draft expansion. 
{\bf Lucas Z.\ Brito}: Software Programming ({\it Quantum.computing} module), Development and Validation, Visualization, Testing.
{\bf Stephen Carr}: Software Programming, Development and Validation, Writing-Original draft expansion.
{\bf Samuele Sanna}: Methodology, Investigation, Supervision, Formal analysis, Resources, Reviewing and Editing. 
{\bf Vesna F. Mitrovi{\'c}}: Conceptualization, Methodology, Supervision, Formal analysis, Visualization, Resources, Writing-Reviewing and Editing.

\section{Acknowledgments}

We thank Prof. Enrico Giampieri and Prof. Sekhar Ramanathan for their helpful advice during  the development of the program. We are grateful to  Jonathan Frassineti for the feedback as the very first user of the program. We also thank Prof.\ Paolo Santini and Prof.\ Alessandro Chiesa for reading the manuscript and providing helpful feedback. 
V. F. M.\ acknowledges support form the U.S.\ National Science Foundation grants OIA-1921199 and DMR-1905532.

\appendix
\section{Form of Different Hamiltonians}
\label{appx:hamiltonians}

The full Hamiltonian of a single-spin nuclear system is given in Eq.\ \ref{NuclearHamiltonian}. Here we expand on terms that are less relevant for physics, but might be useful in other disciplines, along with their secular approximations in the Zeeman dominant regime. To start with, the hyperfine interaction given in Eq.\ \ref{eq:hyperfineFULL} in the secular approximation becomes
\begin{equation}
	\mathcal{H}_{HF} \approx A S_zI_z + B S_zI_x,
\end{equation}
for $A = a_\text{iso} + \hslash b_D(3\cos^2\theta-1), B = 3\hslash b_D\sin\theta\cos\theta$, where $a_\text{iso}$ is the Fermi contact interaction constant. The chemical shift 
term, $\mathcal{H}_{CS}$,   describes the local structure surrounding a nucleus, and thus it is very sample-specific. Its general form is given by
\begin{equation}
	\mathcal{H}_{CS} =
	-\gamma \hslash \mathbf{I} \cdot \bm{\sigma} \cdot  \mathbf{B}_0,
	\label{CSTerm}
\end{equation}
where $\bm{\sigma}$ is the chemical shift tensor, given by 
\begin{equation}
	\bm{\sigma} = \begin{pmatrix}
		\sigma_{xx} & \sigma_{xy} & \sigma_{xz}\\
		\sigma_{yx} & \sigma_{yy} & \sigma_{yz} \\
		\sigma_{zx} & \sigma_{zy} & \sigma_{zz}
	\end{pmatrix}.
\end{equation}
 It depends on the overall electrons around the nuclear site, as well as the orientation of the sample with respect to $\mathbf{B}_0$. The chemical shift in the secular approximation is given by
\begin{equation}
	\mathcal{H}_{CS} \approx -\gamma\hslash\sigma_{zz}(\Theta)B_0,
\end{equation} 
where $\Theta$ is the angle between the molecule and the applied field. The dipolar Hamiltonian $\mathcal{H}_{D}$ is given by 
\begin{equation}
	\mathcal{H}_{D} = \hslash b_D\mathbf{I}^T_1\cdot D \cdot \mathbf{I}_2,
\end{equation}
where $b_D\equiv \frac{\mu_0\gamma_1\gamma_2\hslash}{4\pi r^3_{21}}$ is the dipolar constant, $\mu_0$ is the magnetic constant, $\gamma_1,\gamma_2$ are the gyromagnetic ratio of two interacting spins, and $r_{21}$ is the average distance between the two spins. The quantity $D$ is the tensor that acts between the transpose of the spin operator of the first nucleus $\mathbf{I}^T_1$ and the spin operator of the second nucleus $\mathbf{I}_2$, and is given by 
\begin{equation}
	D = \begin{pmatrix}
		1 - 3\sin^2\theta\cos^2\varphi & 3\sin^2\theta\sin\varphi\cos\varphi & 3\sin\theta\cos\theta\cos\varphi \\
		3\sin^2\theta\sin\varphi\cos\varphi & 1 - 3\sin^2\theta\sin^2\varphi &  3\sin\theta\cos\theta\sin\varphi\\
		3\sin\theta\cos\theta\cos\varphi & 3\sin\theta\cos\theta\sin\varphi & 1 - 3\cos^2\theta
	\end{pmatrix},
\end{equation}
where   $\theta$ is the angle between the distance vector connecting the two spins and the external magnetic field $\mathbf{B}_0$, and $\varphi$ is the azimuthal angle. The dipolar coupling can be approximated in the Zeeman dominant regime for the homonuclear (nuclear-nuclear) \& heteronuclear spins as 
\begin{equation}
	\mathcal{H}_{D1} \approx \hslash b_D \big(\frac{3\cos^2\theta-1}{2}\big)\big[3I_{1z}I_{2z} -\mathbf{I}_1\cdot  \mathbf{I}_2\big],
\end{equation}
and for the heteronuclear spin as
\begin{equation}
	\mathcal{H}_{D2} \approx \hslash b_D \big(3\cos^2\theta-1\big)I_{1z}I_{2z}.
\end{equation}
The J-coupling is given by 
\begin{equation}
	\mathcal{H}_{J} = 2\pi\hslash\mathbf{I}_1\cdot \mathbf{J} \cdot \mathbf{I}_2,
\end{equation}
where $\mathbf{J}$ is the J-coupling tensor, given by 
\begin{equation}
	\mathbf{J} = \begin{pmatrix}
		J_{xx} & J_{xy} & J_{xz}\\
		J_{yx} & J_{yy} & J_{yz} \\
		J_{zx} & J_{zy} & J_{zz}
	\end{pmatrix}.
\end{equation}
In the secular approximation, the J-coupling becomes 
\begin{equation}
	H_J \approx 2\pi \hslash J I_{1z}I_{2z} ,
\end{equation}
where the $J$ constant is much smaller than the difference in the chemical shifts of the two sites \cite{Abragam}.


\bibliographystyle{elsarticle-num}
 \bibliography{DCPaper.bib}

\begin{thebibliography}{10}
\expandafter\ifx\csname url\endcsname\relax
  \def\url#1{\texttt{#1}}\fi
\expandafter\ifx\csname urlprefix\endcsname\relax\def\urlprefix{URL }\fi
\expandafter\ifx\csname href\endcsname\relax
  \def\href#1#2{#2} \def\path#1{#1}\fi

\bibitem{RevModPhys_NMRQC}
L.~M.~K. Vandersypen, I.~L. Chuang,
  \href{https://link.aps.org/doi/10.1103/RevModPhys.76.1037}{{NMR techniques
  for quantum control and computation}}, Rev. Mod. Phys. 76 (2005) 1037--1069.
\newblock \href {http://dx.doi.org/10.1103/RevModPhys.76.1037}
  {\path{doi:10.1103/RevModPhys.76.1037}}.
\newline\urlprefix\url{https://link.aps.org/doi/10.1103/RevModPhys.76.1037}

\bibitem{Ramanathan2004}
C.~Ramanathan, N.~Boulant, Z.~Chen, D.~G. Cory, I.~Chuang, M.~Steffen,
  \href{https://doi.org/10.1007/s11128-004-3668-x}{Nmr quantum information
  processing}, Quantum Information Processing 3~(1) (2004) 15--44.
\newblock \href {http://dx.doi.org/10.1007/s11128-004-3668-x}
  {\path{doi:10.1007/s11128-004-3668-x}}.
\newline\urlprefix\url{https://doi.org/10.1007/s11128-004-3668-x}

\bibitem{Suter20QGate}
S.~S. Hegde, J.~Zhang, D.~Suter,
  \href{https://link.aps.org/doi/10.1103/PhysRevLett.124.220501}{{Efficient
  Quantum Gates for Individual Nuclear Spin Qubits by Indirect Control}}, Phys.
  Rev. Lett. 124 (2020) 220501.
\newblock \href {http://dx.doi.org/10.1103/PhysRevLett.124.220501}
  {\path{doi:10.1103/PhysRevLett.124.220501}}.
\newline\urlprefix\url{https://link.aps.org/doi/10.1103/PhysRevLett.124.220501}

\bibitem{Modi12}
K.~Modi, A.~Brodutch, H.~Cable, T.~Paterek, V.~Vedral,
  \href{https://link.aps.org/doi/10.1103/RevModPhys.84.1655}{{The
  classical-quantum boundary for correlations: Discord and related measures}},
  Rev. Mod. Phys. 84 (2012) 1655--1707.
\newblock \href {http://dx.doi.org/10.1103/RevModPhys.84.1655}
  {\path{doi:10.1103/RevModPhys.84.1655}}.
\newline\urlprefix\url{https://link.aps.org/doi/10.1103/RevModPhys.84.1655}

\bibitem{Kaufmann79}
E.~N. Kaufmann, R.~J. Vianden,
  \href{https://link.aps.org/doi/10.1103/RevModPhys.51.161}{The electric field
  gradient in noncubic metals}, Rev. Mod. Phys. 51 (1979) 161--214.
\newblock \href {http://dx.doi.org/10.1103/RevModPhys.51.161}
  {\path{doi:10.1103/RevModPhys.51.161}}.
\newline\urlprefix\url{https://link.aps.org/doi/10.1103/RevModPhys.51.161}

\bibitem{Halperin86}
W.~P. Halperin,
  \href{https://link.aps.org/doi/10.1103/RevModPhys.58.533}{Quantum size
  effects in metal particles}, Rev. Mod. Phys. 58 (1986) 533--606.
\newblock \href {http://dx.doi.org/10.1103/RevModPhys.58.533}
  {\path{doi:10.1103/RevModPhys.58.533}}.
\newline\urlprefix\url{https://link.aps.org/doi/10.1103/RevModPhys.58.533}

\bibitem{Abragam}
A.~Abragam, Principles of {Nuclear} {Magnetism}, Oxford University Press, 1961.

\bibitem{BlincIC}
R.~Blinc, Magnetic resonance and relaxation in structurally incommensurate
  systems, Phys. Rep. 79 (1981) 331.

\bibitem{Cory97}
D.~G. Cory, A.~F. Fahmy, T.~F. Havel,
  \href{https://www.ncbi.nlm.nih.gov/pmc/articles/PMC19968/}{Ensemble quantum
  computing by {NMR} spectroscopy}, Proceedings of the National Academy of
  Sciences of the United States of America 94~(5) (1997) 1634--1639.
\newline\urlprefix\url{https://www.ncbi.nlm.nih.gov/pmc/articles/PMC19968/}

\bibitem{Chuang1998}
I.~L. Chuang, L.~M.~K. Vandersypen~et al., Experimental realization of a
  quantum algorithm, Nature 393 (1998) 143--146.
\newblock \href {http://dx.doi.org/https://doi.org/10.1038/30181}
  {\path{doi:https://doi.org/10.1038/30181}}.

\bibitem{Jones_1998}
J.~A. Jones, M.~Mosca, \href{http://dx.doi.org/10.1063/1.476739}{Implementation
  of a quantum algorithm on a nuclear magnetic resonance quantum computer}, The
  Journal of Chemical Physics 109~(5) (1998) 1648--1653.
\newblock \href {http://dx.doi.org/10.1063/1.476739}
  {\path{doi:10.1063/1.476739}}.
\newline\urlprefix\url{http://dx.doi.org/10.1063/1.476739}

\bibitem{Jones_1998_2}
J.~A. Jones, M.~Mosca, R.~H. Hansen,
  \href{http://dx.doi.org/10.1038/30687}{Implementation of a quantum search
  algorithm on a quantum computer}, Nature 393~(6683) (1998) 344--346.
\newblock \href {http://dx.doi.org/10.1038/30687} {\path{doi:10.1038/30687}}.
\newline\urlprefix\url{http://dx.doi.org/10.1038/30687}

\bibitem{Vandersypen_2000}
L.~M.~K. Vandersypen, M.~Steffen, G.~Breyta, C.~S. Yannoni, R.~Cleve, I.~L.
  Chuang, \href{http://dx.doi.org/10.1103/PhysRevLett.85.5452}{Experimental
  realization of an order-finding algorithm with an nmr quantum computer},
  Physical Review Letters 85~(25) (2000) 5452--5455.
\newblock \href {http://dx.doi.org/10.1103/physrevlett.85.5452}
  {\path{doi:10.1103/physrevlett.85.5452}}.
\newline\urlprefix\url{http://dx.doi.org/10.1103/PhysRevLett.85.5452}

\bibitem{LONG2001121}
G.~Long, H.~Yan, Y.~Li, C.~Tu, J.~Tao, H.~Chen, M.~Liu, X.~Zhang, J.~Luo,
  L.~Xiao, X.~Zeng,
  \href{https://www.sciencedirect.com/science/article/pii/S0375960101004169}{{Experimental
  NMR realization of a generalized quantum search algorithm}}, Physics Letters
  A 286~(2) (2001) 121--126.
\newblock \href
  {http://dx.doi.org/https://doi.org/10.1016/S0375-9601(01)00416-9}
  {\path{doi:https://doi.org/10.1016/S0375-9601(01)00416-9}}.
\newline\urlprefix\url{https://www.sciencedirect.com/science/article/pii/S0375960101004169}

\bibitem{Sinha01}
N.~Sinha, T.~S. Mahesh, K.~V. Ramanathan, A.~Kumar,
  \href{https://aip.scitation.org/doi/abs/10.1063/1.1346645}{{Toward quantum
  information processing by nuclear magnetic resonance: Pseudopure states and
  logical operations using selective pulses on an oriented spin 3/2 nucleus}},
  The Journal of Chemical Physics 114~(10) (2001) 4415--4420.
\newblock \href
  {http://arxiv.org/abs/https://aip.scitation.org/doi/pdf/10.1063/1.1346645}
  {\path{arXiv:https://aip.scitation.org/doi/pdf/10.1063/1.1346645}}, \href
  {http://dx.doi.org/10.1063/1.1346645} {\path{doi:10.1063/1.1346645}}.
\newline\urlprefix\url{https://aip.scitation.org/doi/abs/10.1063/1.1346645}

\bibitem{Havel02}
T.~F. Havel, D.~G. Cory, S.~Lloyd, N.~Boulant, E.~M. Fortunato, M.~A. Pravia,
  G.~Teklemariam, Y.~S. Weinstein, A.~Bhattacharyya, J.~Hou,
  \href{https://doi.org/10.1119/1.1446857}{{Quantum information processing by
  nuclear magnetic resonance spectroscopy}}, American Journal of Physics 70~(3)
  (2002) 345--362.
\newblock \href {http://arxiv.org/abs/https://doi.org/10.1119/1.1446857}
  {\path{arXiv:https://doi.org/10.1119/1.1446857}}, \href
  {http://dx.doi.org/10.1119/1.1446857} {\path{doi:10.1119/1.1446857}}.
\newline\urlprefix\url{https://doi.org/10.1119/1.1446857}

\bibitem{Xin19}
T.~Xin, B.-X. Wang, K.-R. Li, X.-Y. Kong, S.-J. Wei, T.~Wang, D.~Ruan, G.-L.
  Long, \href{http://stacks.iop.org/1674-1056/27/i=2/a=020308}{{Nuclear
  magnetic resonance for quantum computing: Techniques and recent
  achievements}}, Chinese Physics B 27~(2) (2018) 020308.
\newline\urlprefix\url{http://stacks.iop.org/1674-1056/27/i=2/a=020308}

\bibitem{NMRQIS_Book}
I.~Oliveira, T.~Bonagamba, R.~Sarthour, J.~Freitas, E.~de~Azevedo, {NMR Quantum
  Information Processing}, Elsevier, 2007.

\bibitem{Rao14}
K.~R.~K. Rao, T.~S. Mahesh, A.~Kumar,
  \href{https://link.aps.org/doi/10.1103/PhysRevA.90.012306}{{Efficient
  simulation of unitary operators by combining two numerical algorithms: An NMR
  simulation of the mirror-inversion propagator of an $XY$ spin chain}}, Phys.
  Rev. A 90 (2014) 012306.
\newblock \href {http://dx.doi.org/10.1103/PhysRevA.90.012306}
  {\path{doi:10.1103/PhysRevA.90.012306}}.
\newline\urlprefix\url{https://link.aps.org/doi/10.1103/PhysRevA.90.012306}

\bibitem{Teles18}
J.~Teles, R.~Auccaise, C.~Rivera-Ascona, A.~G. Araujo-Ferreira, J.~P. Andreeta,
  T.~J. Bonagamba, \href{https://doi.org/10.1007/s11128-018-1947-1}{{Spin
  coherent states phenomena probed by quantum state tomography in Zeeman
  perturbed nuclear quadrupole resonance}}, Quantum Information Processing
  17~(7) (2018) 177.
\newblock \href {http://dx.doi.org/10.1007/s11128-018-1947-1}
  {\path{doi:10.1007/s11128-018-1947-1}}.
\newline\urlprefix\url{https://doi.org/10.1007/s11128-018-1947-1}

\bibitem{Lu:2017aa}
D.~Lu, K.~Li, J.~Li, H.~Katiyar, A.~J. Park, G.~Feng, T.~Xin, H.~Li, G.~Long,
  A.~Brodutch, J.~Baugh, B.~Zeng, R.~Laflamme,
  \href{https://doi.org/10.1038/s41534-017-0045-z}{{Enhancing quantum control
  by bootstrapping a quantum processor of 12 qubits}}, npj Quantum Information
  3~(1) (2017) 45.
\newblock \href {http://dx.doi.org/10.1038/s41534-017-0045-z}
  {\path{doi:10.1038/s41534-017-0045-z}}.
\newline\urlprefix\url{https://doi.org/10.1038/s41534-017-0045-z}

\bibitem{Liu2019}
Y.-X. Liu, A.~Ajoy, P.~Cappellaro,
  \href{https://link.aps.org/doi/10.1103/PhysRevLett.122.100501}{Nanoscale
  {Vector} dc {Magnetometry} via {Ancilla}-{Assisted} {Frequency}
  {Up}-{Conversion}}, Physical Review Letters 122~(10) (2019) 100501.
\newblock \href {http://dx.doi.org/10.1103/PhysRevLett.122.100501}
  {\path{doi:10.1103/PhysRevLett.122.100501}}.
\newline\urlprefix\url{https://link.aps.org/doi/10.1103/PhysRevLett.122.100501}

\bibitem{RevModPhys.89.035002}
C.~L. Degen, F.~Reinhard, P.~Cappellaro,
  \href{https://link.aps.org/doi/10.1103/RevModPhys.89.035002}{Quantum
  sensing}, Rev. Mod. Phys. 89 (2017) 035002.
\newblock \href {http://dx.doi.org/10.1103/RevModPhys.89.035002}
  {\path{doi:10.1103/RevModPhys.89.035002}}.
\newline\urlprefix\url{https://link.aps.org/doi/10.1103/RevModPhys.89.035002}

\bibitem{PhysRevLett.102.210502}
P.~Cappellaro, L.~Jiang, J.~S. Hodges, M.~D. Lukin,
  \href{https://link.aps.org/doi/10.1103/PhysRevLett.102.210502}{{Coherence and
  Control of Quantum Registers Based on Electronic Spin in a Nuclear Spin
  Bath}}, Phys. Rev. Lett. 102 (2009) 210502.
\newblock \href {http://dx.doi.org/10.1103/PhysRevLett.102.210502}
  {\path{doi:10.1103/PhysRevLett.102.210502}}.
\newline\urlprefix\url{https://link.aps.org/doi/10.1103/PhysRevLett.102.210502}

\bibitem{PhysRevX.8.021059}
F.~Poggiali, P.~Cappellaro, N.~Fabbri,
  \href{https://link.aps.org/doi/10.1103/PhysRevX.8.021059}{Optimal control for
  one-qubit quantum sensing}, Phys. Rev. X 8 (2018) 021059.
\newblock \href {http://dx.doi.org/10.1103/PhysRevX.8.021059}
  {\path{doi:10.1103/PhysRevX.8.021059}}.
\newline\urlprefix\url{https://link.aps.org/doi/10.1103/PhysRevX.8.021059}

\bibitem{QMetrology20}
H.~Zhou, J.~Choi, S.~Choi, R.~Landig, A.~M. Douglas, J.~Isoya, F.~Jelezko,
  S.~Onoda, H.~Sumiya, P.~Cappellaro, H.~S. Knowles, H.~Park, M.~D. Lukin,
  \href{https://link.aps.org/doi/10.1103/PhysRevX.10.031003}{{Quantum Metrology
  with Strongly Interacting Spin Systems}}, Phys. Rev. X 10 (2020) 031003.
\newblock \href {http://dx.doi.org/10.1103/PhysRevX.10.031003}
  {\path{doi:10.1103/PhysRevX.10.031003}}.
\newline\urlprefix\url{https://link.aps.org/doi/10.1103/PhysRevX.10.031003}

\bibitem{Peng:2021vv}
P.~Peng, C.~Yin, X.~Huang, C.~Ramanathan, P.~Cappellaro,
  \href{https://doi.org/10.1038/s41567-020-01120-z}{Floquet prethermalization
  in dipolar spin chains}, Nature Physics 17~(4) (2021) 444--447.
\newblock \href {http://dx.doi.org/10.1038/s41567-020-01120-z}
  {\path{doi:10.1038/s41567-020-01120-z}}.
\newline\urlprefix\url{https://doi.org/10.1038/s41567-020-01120-z}

\bibitem{CarrMMSpec22}
S.~Carr, I.~K. Nikolov, R.~Cong, A.~D. Maestro, C.~Ramanathan, V.~F.
  Mitrovi{\'c}, {Multi-modal quantum spectroscopy of phase transitions with
  inversion symmetry}, {\it arXiv:2208.10987}\href
  {http://dx.doi.org/10.48550/arXiv.2208.10987}
  {\path{doi:10.48550/arXiv.2208.10987}}.

\bibitem{INikolov20}
I.~K. Nikolov, S.~Carr, A.~G.~D. Maestro, C.~Ramanathan, V.~F. Mitrovi{\'c},
  {Spin squeezing as a probe of emergent quantum orders}, {\it under review at
  J. Phys. Soc. Jpn.}

\bibitem{Johansson2012}
J.~Johansson, P.~Nation, F.~Nori,
  \href{https://www.sciencedirect.com/science/article/pii/S0010465512000835}{Qutip:
  An open-source python framework for the dynamics of open quantum systems},
  Computer Physics Communications 183~(8) (2012) 1760--1772.
\newblock \href {http://dx.doi.org/https://doi.org/10.1016/j.cpc.2012.02.021}
  {\path{doi:https://doi.org/10.1016/j.cpc.2012.02.021}}.
\newline\urlprefix\url{https://www.sciencedirect.com/science/article/pii/S0010465512000835}

\bibitem{QuantumMachines}
{Quantum Machines}, \href{https://www.quantum-machines.co}{System software
  company}, Website.
\newline\urlprefix\url{https://www.quantum-machines.co}

\bibitem{PhysRevLett.127.237201}
L.~V. Pourovskii, D.~F. Mosca, C.~Franchini,
  \href{https://link.aps.org/doi/10.1103/PhysRevLett.127.237201}{Ferro-octupolar
  order and low-energy excitations in ${\mathrm{d}}^{2}$ double perovskites of
  osmium}, Phys. Rev. Lett. 127 (2021) 237201.
\newblock \href {http://dx.doi.org/10.1103/PhysRevLett.127.237201}
  {\path{doi:10.1103/PhysRevLett.127.237201}}.
\newline\urlprefix\url{https://link.aps.org/doi/10.1103/PhysRevLett.127.237201}

\bibitem{Pourovskiie21}
L.~V. Pourovskii, S.~Khmelevskyi,
  \href{https://www.pnas.org/content/118/14/e2025317118}{Hidden order and
  multipolar exchange striction in a correlated f-electron system}, Proceedings
  of the National Academy of Sciences 118~(14).
\newblock \href
  {http://arxiv.org/abs/https://www.pnas.org/content/118/14/e2025317118.full.pdf}
  {\path{arXiv:https://www.pnas.org/content/118/14/e2025317118.full.pdf}},
  \href {http://dx.doi.org/10.1073/pnas.2025317118}
  {\path{doi:10.1073/pnas.2025317118}}.
\newline\urlprefix\url{https://www.pnas.org/content/118/14/e2025317118}

\bibitem{PhysRevLett.127.140604}
G.~Wang, C.~Li, P.~Cappellaro,
  \href{https://link.aps.org/doi/10.1103/PhysRevLett.127.140604}{{Observation
  of Symmetry-Protected Selection Rules in Periodically Driven Quantum
  Systems}}, Phys. Rev. Lett. 127 (2021) 140604.
\newblock \href {http://dx.doi.org/10.1103/PhysRevLett.127.140604}
  {\path{doi:10.1103/PhysRevLett.127.140604}}.
\newline\urlprefix\url{https://link.aps.org/doi/10.1103/PhysRevLett.127.140604}

\bibitem{Carr2021}
S.~Carr, C.~Snider, D.~E. Feldman, C.~Ramanathan, J.~B. Marston, V.~F.
  Mitrovi{\'c}, Signatures of electronic correlations and spin-susceptibility
  anisotropy in nuclear magnetic resonance, arXiv:2110.06811\href
  {http://arxiv.org/abs/2110.06811} {\path{arXiv:2110.06811}}.

\bibitem{PhysRevLett.122.150606}
X.~Turkeshi, T.~Mendes-Santos, G.~Giudici, M.~Dalmonte,
  \href{https://link.aps.org/doi/10.1103/PhysRevLett.122.150606}{{Entanglement-Guided
  Search for Parent Hamiltonians}}, Phys. Rev. Lett. 122 (2019) 150606.
\newblock \href {http://dx.doi.org/10.1103/PhysRevLett.122.150606}
  {\path{doi:10.1103/PhysRevLett.122.150606}}.
\newline\urlprefix\url{https://link.aps.org/doi/10.1103/PhysRevLett.122.150606}

\bibitem{PhysRevLett.124.100605}
W.~Zhu, Z.~Huang, Y.-C. He, X.~Wen,
  \href{https://link.aps.org/doi/10.1103/PhysRevLett.124.100605}{{Entanglement
  Hamiltonian of Many-Body Dynamics in Strongly Correlated Systems}}, Phys.
  Rev. Lett. 124 (2020) 100605.
\newblock \href {http://dx.doi.org/10.1103/PhysRevLett.124.100605}
  {\path{doi:10.1103/PhysRevLett.124.100605}}.
\newline\urlprefix\url{https://link.aps.org/doi/10.1103/PhysRevLett.124.100605}

\bibitem{ALLOUCHE1989171}
A.~Allouche, G.~Pouzard,
  \href{https://www.sciencedirect.com/science/article/pii/0010465589900428}{Computer
  simulation of ft-nmr multiple pulse experiment}, Computer Physics
  Communications 54~(1) (1989) 171--176.
\newblock \href
  {http://dx.doi.org/https://doi.org/10.1016/0010-4655(89)90042-8}
  {\path{doi:https://doi.org/10.1016/0010-4655(89)90042-8}}.
\newline\urlprefix\url{https://www.sciencedirect.com/science/article/pii/0010465589900428}

\bibitem{BAK2000296}
M.~Bak, J.~T. Rasmussen, N.~C. Nielsen,
  \href{https://www.sciencedirect.com/science/article/pii/S1090780700921797}{Simpson:
  A general simulation program for solid-state nmr spectroscopy}, Journal of
  Magnetic Resonance 147~(2) (2000) 296--330.
\newblock \href {http://dx.doi.org/https://doi.org/10.1006/jmre.2000.2179}
  {\path{doi:https://doi.org/10.1006/jmre.2000.2179}}.
\newline\urlprefix\url{https://www.sciencedirect.com/science/article/pii/S1090780700921797}

\bibitem{Eichele}
K.~Eichele,
  \href{http://anorganik.uni-tuebingen.de/klaus/soft/index.php?p=wsolids1/wsolids1}{Wsolids1
  ver.\ 1.21.7}.
\newline\urlprefix\url{http://anorganik.uni-tuebingen.de/klaus/soft/index.php?p=wsolids1/wsolids1}

\bibitem{reich2002}
H.~J. Reich,
  \href{https://www2.chem.wisc.edu/areas/reich/plt/windnmr.htm}{Windnmr-pro},
  Windows program (Feb. 2002).
\newline\urlprefix\url{https://www2.chem.wisc.edu/areas/reich/plt/windnmr.htm}

\bibitem{Possa}
D.~Possa, A.~C. Gaudio, J.~C.~C. Freitas, Numerical simulation of {NQR/NMR}:
  {Applications} in quantum computing, Journal of Magnetic Resonance 209~(2)
  (2011) 250--260.

\bibitem{Bengs2018}
C.~Bengs, M.~H. Levitt,
  \href{https://analyticalsciencejournals.onlinelibrary.wiley.com/doi/abs/10.1002/mrc.4642}{Spindynamica:
  Symbolic and numerical magnetic resonance in a mathematica environment},
  Magnetic Resonance in Chemistry 56~(6) (2018) 374--414.
\newblock \href
  {http://arxiv.org/abs/https://analyticalsciencejournals.onlinelibrary.wiley.com/doi/pdf/10.1002/mrc.4642}
  {\path{arXiv:https://analyticalsciencejournals.onlinelibrary.wiley.com/doi/pdf/10.1002/mrc.4642}},
  \href {http://dx.doi.org/https://doi.org/10.1002/mrc.4642}
  {\path{doi:https://doi.org/10.1002/mrc.4642}}.
\newline\urlprefix\url{https://analyticalsciencejournals.onlinelibrary.wiley.com/doi/abs/10.1002/mrc.4642}

\bibitem{Hogben2011}
H.~Hogben, M.~Krzystyniak, G.~Charnock, P.~Hore, I.~Kuprov,
  \href{https://www.sciencedirect.com/science/article/pii/S1090780710003575}{Spinach
  -- a software library for simulation of spin dynamics in large spin systems},
  Journal of Magnetic Resonance 208~(2) (2011) 179--194.
\newblock \href {http://dx.doi.org/https://doi.org/10.1016/j.jmr.2010.11.008}
  {\path{doi:https://doi.org/10.1016/j.jmr.2010.11.008}}.
\newline\urlprefix\url{https://www.sciencedirect.com/science/article/pii/S1090780710003575}

\bibitem{Veshtort2006}
M.~Veshtort, R.~G. Griffin,
  \href{https://www.sciencedirect.com/science/article/pii/S1090780705002442}{Spinevolution:
  A powerful tool for the simulation of solid and liquid state nmr
  experiments}, Journal of Magnetic Resonance 178~(2) (2006) 248--282.
\newblock \href {http://dx.doi.org/https://doi.org/10.1016/j.jmr.2005.07.018}
  {\path{doi:https://doi.org/10.1016/j.jmr.2005.07.018}}.
\newline\urlprefix\url{https://www.sciencedirect.com/science/article/pii/S1090780705002442}

\bibitem{perchNMR}
{PERCH Solutions Ltd.}, \href{http://new.perchsolutions.com/}{Perch nmr
  software}.
\newline\urlprefix\url{http://new.perchsolutions.com/}

\bibitem{PERRAS201236}
F.~A. Perras, C.~M. Widdifield, D.~L. Bryce,
  \href{https://www.sciencedirect.com/science/article/pii/S0926204012000586}{{QUEST
  -- Quadrupolar Exact Software: A fast graphical program for the exact
  simulation of NMR and NQR spectra for quadrupolar nuclei}}, Solid State
  Nuclear Magnetic Resonance 45-46 (2012) 36--44.
\newblock \href {http://dx.doi.org/https://doi.org/10.1016/j.ssnmr.2012.05.002}
  {\path{doi:https://doi.org/10.1016/j.ssnmr.2012.05.002}}.
\newline\urlprefix\url{https://www.sciencedirect.com/science/article/pii/S0926204012000586}

\bibitem{Binev2007}
Y.~Binev, M.~M.~B. Marques, J.~Aires-de Sousa,
  \href{https://doi.org/10.1021/ci700172n}{Prediction of 1h nmr coupling
  constants with associative neural networks trained for chemical shifts}, J.
  Chem. Inf. Model. 47~(6) (2007) 2089--2097.
\newblock \href {http://dx.doi.org/10.1021/ci700172n}
  {\path{doi:10.1021/ci700172n}}.
\newline\urlprefix\url{https://doi.org/10.1021/ci700172n}

\bibitem{Claridge2009}
T.~Claridge, \href{https://doi.org/10.1021/ci900090d}{Software review of mnova:
  Nmr data processing, analysis, and prediction software}, J. Chem. Inf. Model.
  49~(4) (2009) 1136--1137.
\newblock \href {http://dx.doi.org/10.1021/ci900090d}
  {\path{doi:10.1021/ci900090d}}.
\newline\urlprefix\url{https://doi.org/10.1021/ci900090d}

\bibitem{Schwieters2001}
C.~D. Schwieters, G.~M. Clore,
  \href{https://www.sciencedirect.com/science/article/pii/S1090780701923006}{The
  vmd-xplor visualization package for nmr structure refinement}, Journal of
  Magnetic Resonance 149~(2) (2001) 239--244.
\newblock \href {http://dx.doi.org/10.1006/jmre.2001.2300}
  {\path{doi:10.1006/jmre.2001.2300}}.
\newline\urlprefix\url{https://www.sciencedirect.com/science/article/pii/S1090780701923006}

\bibitem{PULSEE}
D.~Candoli, \href{https://github.com/vemiBGH/PULSEE}{{PULSEE} ({Program} for
  the {simULation} of {nuclear} {Spin} {Ensemble} {Evolution}} (2021).
\newline\urlprefix\url{https://github.com/vemiBGH/PULSEE}

\bibitem{Snider2022}
C.~Snider, S.~Carr, D.~E. Feldman, C.~Ramanathan, J.~B. Marston, V.~F.
  Mitrovi{\'c}, Simulation of spin echo and dynamics of interacting nuclear
  spins, To be submitted to Computer Physics Communications.

\bibitem{Lu17}
L.~Lu, M.~Song, W.~Liu, A.~P. Reyes, P.~Kuhns, H.~O. Lee, I.~R. Fisher, V.~F.
  Mitrovi{\'c}, {Magnetism and local symmetry breaking in a Mott insulator with
  strong spin orbit interactions}, Nature Communications 8~(14407).

\bibitem{RongOrbit19}
R.~Cong, R.~Nanguneri, B.~Rubenstein, V.~F. Mitrovi\ifmmode~\acute{c}\else
  \'{c}\fi{},
  \href{https://link.aps.org/doi/10.1103/PhysRevB.100.245141}{Evidence from
  first-principles calculations for orbital ordering in
  ${\mathrm{ba}}_{2}{\mathrm{naoso}}_{6}$: A mott insulator with strong
  spin-orbit coupling}, Phys. Rev. B 100 (2019) 245141.
\newblock \href {http://dx.doi.org/10.1103/PhysRevB.100.245141}
  {\path{doi:10.1103/PhysRevB.100.245141}}.
\newline\urlprefix\url{https://link.aps.org/doi/10.1103/PhysRevB.100.245141}

\bibitem{SuterNV20}
J.~Zhang, S.~S. Hegde, D.~Suter,
  \href{https://link.aps.org/doi/10.1103/PhysRevLett.125.030501}{Efficient
  implementation of a quantum algorithm in a single nitrogen-vacancy center of
  diamond}, Phys. Rev. Lett. 125 (2020) 030501.
\newblock \href {http://dx.doi.org/10.1103/PhysRevLett.125.030501}
  {\path{doi:10.1103/PhysRevLett.125.030501}}.
\newline\urlprefix\url{https://link.aps.org/doi/10.1103/PhysRevLett.125.030501}

\bibitem{Blanes}
S.~Blanes, F.~Casas, J.~A. Oteo, J.~Ros, A pedagogical approach to the {Magnus}
  expansion, Eur. J. Phys. 31 (2010) 907--918.

\bibitem{Slichter}
C.~P. Slichter, Principles of {Magnetic} {Resonance}, Springer-Verlag, 1990.

\bibitem{Hore}
P.~J. Hore, J.~A. Jones, S.~Wimperis, NMR: {The} {Toolkit}. {How} {Pulse}
  {Sequences} {Work}, Oxford, 2015.

\bibitem{McCoy1989}
M.~A. McCoy, R.~R. Ernst,
  \href{https://www.sciencedirect.com/science/article/pii/0009261489875372}{Nuclear
  spin noise at room temperature}, Chemical Physics Letters 159~(5) (1989)
  587--593.
\newblock \href {http://dx.doi.org/10.1016/0009-2614(89)87537-2}
  {\path{doi:10.1016/0009-2614(89)87537-2}}.
\newline\urlprefix\url{https://www.sciencedirect.com/science/article/pii/0009261489875372}

\bibitem{D.K.1998}
D.-K. Yang, J.~E. Atkins, C.~C. Lester, D.~B. Zax,
  \href{https://doi.org/10.1080/00268976.2011.9720930}{New developments in
  nuclear magnetic resonance using noise spectroscopy}, Molecular Physics
  95~(5) (1998) 747--757, publisher: Taylor \& Francis \_eprint:
  https://doi.org/10.1080/00268976.2011.9720930.
\newblock \href {http://dx.doi.org/10.1080/00268976.2011.9720930}
  {\path{doi:10.1080/00268976.2011.9720930}}.
\newline\urlprefix\url{https://doi.org/10.1080/00268976.2011.9720930}

\bibitem{Ferrand2015}
G.~Ferrand, G.~Huber, M.~Luong, H.~Desvaux,
  \href{https://doi.org/10.1063/1.4929783}{Nuclear spin noise in nmr
  revisited}, J. Chem. Phys. 143~(9) (2015) 094201.
\newblock \href {http://dx.doi.org/10.1063/1.4929783}
  {\path{doi:10.1063/1.4929783}}.
\newline\urlprefix\url{https://doi.org/10.1063/1.4929783}

\bibitem{Ajoy2019}
A.~Ajoy, B.~Safvati, R.~Nazaryan, J.~T. Oon, B.~Han, P.~Raghavan, R.~Nirodi,
  A.~Aguilar, K.~Liu, X.~Cai, X.~Lv, E.~Druga, C.~Ramanathan, J.~A. Reimer,
  C.~A. Meriles, D.~Suter, A.~Pines,
  \href{https://www.nature.com/articles/s41467-019-13042-3}{Hyperpolarized
  relaxometry based nuclear {T1} noise spectroscopy in diamond}, Nature
  Communications 10~(1) (2019) 5160, number: 1 Publisher: Nature Publishing
  Group.
\newblock \href {http://dx.doi.org/10.1038/s41467-019-13042-3}
  {\path{doi:10.1038/s41467-019-13042-3}}.
\newline\urlprefix\url{https://www.nature.com/articles/s41467-019-13042-3}

\bibitem{Sung2021}
Y.~Sung, A.~Veps{\"a}l{\"a}inen, J.~Braum{\"u}ller, F.~Yan, J.~I.-J. Wang,
  M.~Kjaergaard, R.~Winik, P.~Krantz, A.~Bengtsson, A.~J. Melville, B.~M.
  Niedzielski, M.~E. Schwartz, D.~K. Kim, J.~L. Yoder, T.~P. Orlando,
  S.~Gustavsson, W.~D. Oliver,
  \href{https://www.nature.com/articles/s41467-021-21098-3}{Multi-level quantum
  noise spectroscopy}, Nature Communications 12~(1) (2021) 967, number: 1
  Publisher: Nature Publishing Group.
\newblock \href {http://dx.doi.org/10.1038/s41467-021-21098-3}
  {\path{doi:10.1038/s41467-021-21098-3}}.
\newline\urlprefix\url{https://www.nature.com/articles/s41467-021-21098-3}

\bibitem{Weber1960}
M.~J. Weber, E.~L. Hahn,
  \href{https://link.aps.org/doi/10.1103/PhysRev.120.365}{Selective spin
  excitation and relaxation in nuclear quadrupole resonance}, Phys. Rev. 120
  (1960) 365--375.
\newblock \href {http://dx.doi.org/10.1103/PhysRev.120.365}
  {\path{doi:10.1103/PhysRev.120.365}}.
\newline\urlprefix\url{https://link.aps.org/doi/10.1103/PhysRev.120.365}

\bibitem{LEE2001355}
Y.~Lee, H.~Robert, D.~Lathrop,
  \href{https://www.sciencedirect.com/science/article/pii/S1090780700922481}{{Circular
  Polarization Excitation and Detection in $^{14}$N NQR}}, Journal of Magnetic
  Resonance 148~(2) (2001) 355--362.
\newblock \href {http://dx.doi.org/https://doi.org/10.1006/jmre.2000.2248}
  {\path{doi:https://doi.org/10.1006/jmre.2000.2248}}.
\newline\urlprefix\url{https://www.sciencedirect.com/science/article/pii/S1090780700922481}

\bibitem{MILLER2001228}
J.~Miller, B.~Suits, A.~Garroway,
  \href{https://www.sciencedirect.com/science/article/pii/S1090780701923663}{Circularly
  polarized rf magnetic fields for spin-1 nqr}, Journal of Magnetic Resonance
  151~(2) (2001) 228--234.
\newblock \href {http://dx.doi.org/https://doi.org/10.1006/jmre.2001.2366}
  {\path{doi:https://doi.org/10.1006/jmre.2001.2366}}.
\newline\urlprefix\url{https://www.sciencedirect.com/science/article/pii/S1090780701923663}

\bibitem{Das}
T.~P. Das, E.~L. Hahn, {Nuclear} {Quadrupole} {Resonance} {Spectroscopy}.

\bibitem{Cory}
D.~Cory, R.~Laflamme, E.~Knill, L.~Viola, T.~Havel, N.~Boulant, G.~Boutis,
  E.~Fortunato, S.~Lloyd, R.~Martinez, C.~Negrevergne, M.~Pravia, Y.~Sharf,
  G.~Teklemariam, Y.~Weinstein, W.~Zurek, Nmr based quantum information
  processing: Achievements and prospects, Fortschritte der Physik 48~(9‐11)
  (2000) 875--907.
\newblock \href
  {http://dx.doi.org/https://doi.org/10.1002/1521-3978(200009)48:9/11<875::AID-PROP875>3.0.CO;2-V}
  {\path{doi:https://doi.org/10.1002/1521-3978(200009)48:9/11<875::AID-PROP875>3.0.CO;2-V}}.

\bibitem{Murali}
K.~V. R.~M. Murali, N.~Sinha, T.~S. Mahesh, M.~H. Levitt, K.~V. Ramanathan,
  A.~Kumar,
  \href{https://link.aps.org/doi/10.1103/PhysRevA.66.022313}{Quantum-information
  processing by nuclear magnetic resonance: Experimental implementation of
  half-adder and subtractor operations using an oriented spin-7/2 system},
  Phys. Rev. A 66 (2002) 022313.
\newblock \href {http://dx.doi.org/10.1103/PhysRevA.66.022313}
  {\path{doi:10.1103/PhysRevA.66.022313}}.
\newline\urlprefix\url{https://link.aps.org/doi/10.1103/PhysRevA.66.022313}

\bibitem{Kampermann}
H.~Kampermann, W.~S. Veeman,
  \href{https://doi.org/10.1063/1.1904595}{Characterization of quantum
  algorithms by quantum process tomography using quadrupolar spins in
  solid-state nuclear magnetic resonance}, The Journal of Chemical Physics
  122~(21) (2005) 214108.
\newblock \href {http://arxiv.org/abs/https://doi.org/10.1063/1.1904595}
  {\path{arXiv:https://doi.org/10.1063/1.1904595}}, \href
  {http://dx.doi.org/10.1063/1.1904595} {\path{doi:10.1063/1.1904595}}.
\newline\urlprefix\url{https://doi.org/10.1063/1.1904595}

\bibitem{Oliveira2012}
R.~M. Serra, I.~S. Oliveira, Nuclear magnetic resonance quantum information
  processing, Phil. Trans. R. Soc. A 370 (2012) 4615--4619.
\newblock \href {http://dx.doi.org/https://doi.org/10.1098/rsta.2012.0332}
  {\path{doi:https://doi.org/10.1098/rsta.2012.0332}}.

\bibitem{Jones1998}
J.~A. Jones, R.~H. Hansen, M.~Mosca,
  \href{http://arxiv.org/abs/quant-ph/9805070}{Quantum {Logic} {Gates} and
  {Nuclear} {Magnetic} {Resonance} {Pulse} {Sequences}}, Journal of Magnetic
  Resonance 135~(2) (1998) 353--360, arXiv: quant-ph/9805070.
\newblock \href {http://dx.doi.org/10.1006/jmre.1998.1606}
  {\path{doi:10.1006/jmre.1998.1606}}.
\newline\urlprefix\url{http://arxiv.org/abs/quant-ph/9805070}

\bibitem{Price1999}
M.~Price, S.~Somaroo, C.~Tseng, J.~Gore, A.~Fahmy, T.~Havel, D.~Cory,
  \href{https://linkinghub.elsevier.com/retrieve/pii/S1090780799918517}{Construction
  and {Implementation} of {NMR} {Quantum} {Logic} {Gates} for {Two} {Spin}
  {Systems}}, Journal of Magnetic Resonance 140~(2) (1999) 371--378.
\newblock \href {http://dx.doi.org/10.1006/jmre.1999.1851}
  {\path{doi:10.1006/jmre.1999.1851}}.
\newline\urlprefix\url{https://linkinghub.elsevier.com/retrieve/pii/S1090780799918517}

\bibitem{Dorai}
K.~Dorai, Arvind, A.~Kumar,
  \href{https://link.aps.org/doi/10.1103/PhysRevA.61.042306}{Implementing
  quantum-logic operations, pseudopure states, and the deutsch-jozsa algorithm
  using noncommuting selective pulses in nmr}, Phys. Rev. A 61 (2000) 042306.
\newblock \href {http://dx.doi.org/10.1103/PhysRevA.61.042306}
  {\path{doi:10.1103/PhysRevA.61.042306}}.
\newline\urlprefix\url{https://link.aps.org/doi/10.1103/PhysRevA.61.042306}

\bibitem{Guelec2005}
A.~G{\"u}le{\c c}, S.~Bah{\c c}eli,
  \href{http://przyrbwn.icm.edu.pl/APP/PDF/107/a107z609.pdf}{Quantum {Gates}
  {Based} on {Polarization} {Transfer} {NMR} {Spectroscopy}}, Acta Physica
  Polonica A 107~(6) (2005) 983--989.
\newblock \href {http://dx.doi.org/10.12693/APhysPolA.107.983}
  {\path{doi:10.12693/APhysPolA.107.983}}.
\newline\urlprefix\url{http://przyrbwn.icm.edu.pl/APP/PDF/107/a107z609.pdf}

\bibitem{Teles2012}
J.~Teles, E.~R. DeAzevedo, J.~C.~C. Freitas, R.~S. Sarthour, I.~S. Oliveira,
  T.~J. Bonagamba,
  \href{https://royalsocietypublishing.org/doi/10.1098/rsta.2011.0365}{Quantum
  information processing by nuclear magnetic resonance on quadrupolar nuclei},
  Philosophical Transactions of the Royal Society A: Mathematical, Physical and
  Engineering Sciences 370~(1976) (2012) 4770--4793.
\newblock \href {http://dx.doi.org/10.1098/rsta.2011.0365}
  {\path{doi:10.1098/rsta.2011.0365}}.
\newline\urlprefix\url{https://royalsocietypublishing.org/doi/10.1098/rsta.2011.0365}

\bibitem{Tan2012}
Y.-P. Tan, X.-F. Nie, J.~Li, H.-W. Chen, X.-Y. Zhou, X.-H. Peng, J.-F. Du,
  \href{https://doi.org/10.1088/0256-307x/29/12/127601}{Preparing pseudo-pure
  states in a quadrupolar spin system using optimal control}, Chinese Physics
  Letters 29~(12) (2012) 127601.
\newblock \href {http://dx.doi.org/10.1088/0256-307x/29/12/127601}
  {\path{doi:10.1088/0256-307x/29/12/127601}}.
\newline\urlprefix\url{https://doi.org/10.1088/0256-307x/29/12/127601}

\bibitem{Teles2015}
J.~Teles, C.~Rivera-Ascona, R.~S. Polli, R.~Oliveira-Silva, E.~L.~G. Vidoto,
  J.~P. Andreeta, T.~J. Bonagamba,
  \href{http://link.springer.com/10.1007/s11128-015-0967-3}{Experimental
  implementation of quantum information processing by {Zeeman}-perturbed
  nuclear quadrupole resonance}, Quantum Information Processing 14~(6) (2015)
  1889--1906.
\newblock \href {http://dx.doi.org/10.1007/s11128-015-0967-3}
  {\path{doi:10.1007/s11128-015-0967-3}}.
\newline\urlprefix\url{http://link.springer.com/10.1007/s11128-015-0967-3}

\bibitem{Wolfowicz2016}
G.~Wolfowicz, J.~J. Morton,
  \href{http://doi.wiley.com/10.1002/9780470034590.emrstm1521}{Pulse
  {Techniques} for {Quantum} {Information} {Processing}}, in: R.~K. Harris,
  R.~L. Wasylishen (Eds.), {eMagRes}, John Wiley \& Sons, Ltd, Chichester, UK,
  2016, pp. 1515--1528.
\newblock \href {http://dx.doi.org/10.1002/9780470034590.emrstm1521}
  {\path{doi:10.1002/9780470034590.emrstm1521}}.
\newline\urlprefix\url{http://doi.wiley.com/10.1002/9780470034590.emrstm1521}

\bibitem{Min2022}
J.~Min, W.~Teng, W.~Blanchard~John, F.~Guanru, P.~Xinhua, B.~Dmitry,
  \href{https://doi.org/10.1126/sciadv.aar6327}{Experimental benchmarking of
  quantum control in zero-field nuclear magnetic resonance}, Science Advances
  4~(6) (2022) eaar6327.
\newblock \href {http://dx.doi.org/10.1126/sciadv.aar6327}
  {\path{doi:10.1126/sciadv.aar6327}}.
\newline\urlprefix\url{https://doi.org/10.1126/sciadv.aar6327}

\bibitem{Knill98}
E.~Knill, I.~Chuang, R.~Laflamme,
  \href{https://link.aps.org/doi/10.1103/PhysRevA.57.3348}{{Effective pure
  states for bulk quantum computation}}, Phys. Rev. A 57 (1998) 3348--3363.
\newblock \href {http://dx.doi.org/10.1103/PhysRevA.57.3348}
  {\path{doi:10.1103/PhysRevA.57.3348}}.
\newline\urlprefix\url{https://link.aps.org/doi/10.1103/PhysRevA.57.3348}

\bibitem{Feldman2008}
E.~B. Fel'dman, A.~N. Pyrkov,
  \href{http://link.springer.com/10.1134/S0021364008180124}{Evolution of spin
  entanglement and an entanglement witness in multiple-quantum {NMR}
  experiments}, JETP Letters 88~(6) (2008) 398--401.
\newblock \href {http://dx.doi.org/10.1134/S0021364008180124}
  {\path{doi:10.1134/S0021364008180124}}.
\newline\urlprefix\url{http://link.springer.com/10.1134/S0021364008180124}

\bibitem{Gerasev2018}
S.~A. Gerasev, A.~V. Fedorova, E.~B. Fel'dman, E.~I. Kuznetsova,
  \href{http://arxiv.org/abs/1802.09042}{Theoretical investigations of quantum
  correlations in {NMR} multiple-pulse spin-locking experiments}, Quantum
  Information Processing 17~(4) (2018) 72, arXiv: 1802.09042.
\newblock \href {http://dx.doi.org/10.1007/s11128-018-1841-x}
  {\path{doi:10.1007/s11128-018-1841-x}}.
\newline\urlprefix\url{http://arxiv.org/abs/1802.09042}

\bibitem{Gaerttner2018}
M.~G{\"a}rttner, P.~Hauke, A.~M. Rey,
  \href{http://arxiv.org/abs/1706.01616}{Relating out-of-time-order
  correlations to entanglement via multiple-quantum coherences}, Physical
  Review Letters 120~(4) (2018) 040402, arXiv: 1706.01616.
\newblock \href {http://dx.doi.org/10.1103/PhysRevLett.120.040402}
  {\path{doi:10.1103/PhysRevLett.120.040402}}.
\newline\urlprefix\url{http://arxiv.org/abs/1706.01616}

\bibitem{PhysRevX.12.011016}
Y.~Suzuki, K.~Wakamatsu, J.~Ibuka, H.~Oike, T.~Fujii, K.~Miyagawa,
  H.~Taniguchi, K.~Kanoda,
  \href{https://link.aps.org/doi/10.1103/PhysRevX.12.011016}{{Mott-Driven
  BEC-BCS Crossover in a Doped Spin Liquid Candidate
  $\kappa$-BEDT-TTF$_{4}$Hg$_{2.89}$Br$_{8}$ }}, Phys. Rev. X 12 (2022) 011016.
\newblock \href {http://dx.doi.org/10.1103/PhysRevX.12.011016}
  {\path{doi:10.1103/PhysRevX.12.011016}}.
\newline\urlprefix\url{https://link.aps.org/doi/10.1103/PhysRevX.12.011016}

\bibitem{Ticozzi2017}
F.~Ticozzi, L.~Viola, \href{http://arxiv.org/abs/1704.01486}{Quantum and
  classical resources for unitary design of open-system evolutions}, Quantum
  Science and Technology 2~(3) (2017) 034001, arXiv: 1704.01486.
\newblock \href {http://dx.doi.org/10.1088/2058-9565/aa722a}
  {\path{doi:10.1088/2058-9565/aa722a}}.
\newline\urlprefix\url{http://arxiv.org/abs/1704.01486}

\bibitem{Ajoy2018}
A.~Ajoy, R.~Nazaryan, K.~Liu, X.~Lv, B.~Safvati, G.~Wang, E.~Druga, J.~A.
  Reimer, D.~Suter, C.~Ramanathan, C.~A. Meriles, A.~Pines,
  \href{http://www.pnas.org/lookup/doi/10.1073/pnas.1807125115}{Enhanced
  dynamic nuclear polarization via swept microwave frequency combs},
  Proceedings of the National Academy of Sciences 115~(42) (2018) 10576--10581.
\newblock \href {http://dx.doi.org/10.1073/pnas.1807125115}
  {\path{doi:10.1073/pnas.1807125115}}.
\newline\urlprefix\url{http://www.pnas.org/lookup/doi/10.1073/pnas.1807125115}

\bibitem{ChenCurr17}
J.~Tian, S.~Hong, I.~Miotkowski, S.~Datta, Y.~P. Chen,
  \href{https://www.science.org/doi/abs/10.1126/sciadv.1602531}{{Observation of
  current-induced, long-lived persistent spin polarization in a topological
  insulator: A rechargeable spin battery}}, Science Advances 3~(4) (2017)
  e1602531.
\newblock \href
  {http://arxiv.org/abs/https://www.science.org/doi/pdf/10.1126/sciadv.1602531}
  {\path{arXiv:https://www.science.org/doi/pdf/10.1126/sciadv.1602531}}, \href
  {http://dx.doi.org/10.1126/sciadv.1602531}
  {\path{doi:10.1126/sciadv.1602531}}.
\newline\urlprefix\url{https://www.science.org/doi/abs/10.1126/sciadv.1602531}

\bibitem{PhysRevA.105.022428}
M.~Ahsan, S.~A.~Z. Naqvi, H.~Anwer,
  \href{https://link.aps.org/doi/10.1103/PhysRevA.105.022428}{Quantum circuit
  engineering for correcting coherent noise}, Phys. Rev. A 105 (2022) 022428.
\newblock \href {http://dx.doi.org/10.1103/PhysRevA.105.022428}
  {\path{doi:10.1103/PhysRevA.105.022428}}.
\newline\urlprefix\url{https://link.aps.org/doi/10.1103/PhysRevA.105.022428}

\bibitem{LaflammeNNhf11}
Y.~Zhang, C.~A. Ryan, R.~Laflamme, J.~Baugh,
  \href{https://link.aps.org/doi/10.1103/PhysRevLett.107.170503}{Coherent
  control of two nuclear spins using the anisotropic hyperfine interaction},
  Phys. Rev. Lett. 107 (2011) 170503.
\newblock \href {http://dx.doi.org/10.1103/PhysRevLett.107.170503}
  {\path{doi:10.1103/PhysRevLett.107.170503}}.
\newline\urlprefix\url{https://link.aps.org/doi/10.1103/PhysRevLett.107.170503}

\bibitem{Koutroulakis10}
G.~Koutroulakis, M.~D. Stewart, V.~F. Mitrovi\ifmmode~\acute{c}\else
  \'{c}\fi{}, M.~Horvati\ifmmode~\acute{c}\else \'{c}\fi{}, C.~Berthier,
  G.~Lapertot, J.~Flouquet,
  \href{http://link.aps.org/doi/10.1103/PhysRevLett.104.087001}{{Field
  Evolution of Coexisting Superconducting and Magnetic Orders in
  $\mathrm{CeCoIn_{5}}$}}, Phys. Rev. Lett. 104 (2010) 087001.
\newblock \href {http://dx.doi.org/10.1103/PhysRevLett.104.087001}
  {\path{doi:10.1103/PhysRevLett.104.087001}}.
\newline\urlprefix\url{http://link.aps.org/doi/10.1103/PhysRevLett.104.087001}

\bibitem{Radcliffe1971}
J.~M. Radcliffe,
  \href{https://iopscience.iop.org/article/10.1088/0305-4470/4/3/009}{Some
  properties of coherent spin states}, Journal of Physics A: General Physics
  4~(3) (1971) 313--323.
\newblock \href {http://dx.doi.org/10.1088/0305-4470/4/3/009}
  {\path{doi:10.1088/0305-4470/4/3/009}}.
\newline\urlprefix\url{https://iopscience.iop.org/article/10.1088/0305-4470/4/3/009}

\bibitem{AuccaiseEstrada2013}
R.~Auccaise~Estrada, E.~R. de~Azevedo, E.~I. Duzzioni, T.~J. Bonagamba, M.~H.
  Youssef~Moussa,
  \href{http://link.springer.com/10.1140/epjd/e2013-30689-1}{Spin coherent
  states in {NMR} quadrupolar system: experimental and theoretical
  applications}, The European Physical Journal D 67~(6) (2013) 127.
\newblock \href {http://dx.doi.org/10.1140/epjd/e2013-30689-1}
  {\path{doi:10.1140/epjd/e2013-30689-1}}.
\newline\urlprefix\url{http://link.springer.com/10.1140/epjd/e2013-30689-1}

\bibitem{Auccaise2015}
R.~Auccaise, A.~Araujo-Ferreira, R.~Sarthour, I.~Oliveira, T.~Bonagamba,
  I.~Roditi,
  \href{https://link.aps.org/doi/10.1103/PhysRevLett.114.043604}{Spin
  {Squeezing} in a {Quadrupolar} {Nuclei} {NMR} {System}}, Physical Review
  Letters 114~(4) (2015) 043604.
\newblock \href {http://dx.doi.org/10.1103/PhysRevLett.114.043604}
  {\path{doi:10.1103/PhysRevLett.114.043604}}.
\newline\urlprefix\url{https://link.aps.org/doi/10.1103/PhysRevLett.114.043604}

\bibitem{Sinha2003}
S.~Sinha, J.~Emerson, N.~Boulant, E.~M. Fortunato, T.~F. Havel, D.~G. Cory,
  \href{https://doi.org/10.1023/B:QINP.0000042202.87144.cb}{Experimental
  simulation of spin squeezing by nuclear magnetic resonance}, Quantum
  Information Processing 2~(6) (2003) 433--448.
\newblock \href {http://dx.doi.org/10.1023/B:QINP.0000042202.87144.cb}
  {\path{doi:10.1023/B:QINP.0000042202.87144.cb}}.
\newline\urlprefix\url{https://doi.org/10.1023/B:QINP.0000042202.87144.cb}

\bibitem{Fortunato2002}
E.~M. Fortunato, M.~A. Pravia, N.~Boulant, G.~Teklemariam, T.~F. Havel, D.~G.
  Cory, \href{https://aip.scitation.org/doi/10.1063/1.1465412}{Design of
  strongly modulating pulses to implement precise effective {Hamiltonians} for
  quantum information processing}, The Journal of Chemical Physics 116~(17)
  (2002) 7599--7606, publisher: American Institute of Physics.
\newblock \href {http://dx.doi.org/10.1063/1.1465412}
  {\path{doi:10.1063/1.1465412}}.
\newline\urlprefix\url{https://aip.scitation.org/doi/10.1063/1.1465412}

\bibitem{Gaikwad2018}
A.~Gaikwad, D.~Rehal, A.~Singh, {Arvind}, K.~Dorai,
  \href{https://link.aps.org/doi/10.1103/PhysRevA.97.022311}{Experimental
  demonstration of selective quantum process tomography on an {NMR} quantum
  information processor}, Physical Review A 97~(2) (2018) 022311, publisher:
  American Physical Society.
\newblock \href {http://dx.doi.org/10.1103/PhysRevA.97.022311}
  {\path{doi:10.1103/PhysRevA.97.022311}}.
\newline\urlprefix\url{https://link.aps.org/doi/10.1103/PhysRevA.97.022311}

\bibitem{Baaquie2013}
B.~E. Baaquie, \href{https://doi.org/10.1007/978-1-4614-6224-8_6}{The
  Theoretical Foundations of Quantum Mechanics}, Springer, New York, NY, 2013,
  Ch. Density Matrix: Entangled States, pp. 93--113.
\newline\urlprefix\url{https://doi.org/10.1007/978-1-4614-6224-8_6}

\bibitem{NikolovAPS21}
I.~Nikolov, A.~D. Maestro, C.~Ramanathan, V.~F. Mitrovi{\'c}, {Spin Squeezing
  as a Probe of Quantum Materials}, APS - March Meeting Bulletin - V44.00014
  (2021).

\end{thebibliography}


\begin{thebibliography}{0}
\bibitem{1} F. A. Perras, C. M. Widdifield, and D. L. Bryce, ``{\it QUEST - Quadrupolar Exact Software: A fast graphical program for the exact simulation of NMR and NQR spectra for quadrupolar nuclei},'' Solid State Nuclear Magnetic Resonance, vol. 45-46, pp. 36-44, (2012).      
\bibitem{2} D. Possa, A. C. Gaudio, and J. C. C. Freitas, ``Numerical simulation of NQR/NMR: Applications in quantum computing,'' Journal of Magnetic Resonance, vol. 209, pp. 250-260,  (2011).      
\end{thebibliography}








\end{document}